\UseRawInputEncoding
\documentclass[a4paper,11pt]{article}
\pdfoutput=1 % if your are submitting a pdflatex (i.e. if you have
\usepackage{jheppub} % for details on the use of the package, please
\usepackage{hyperref}
\usepackage{cleveref}
\def\lsim{\raise0.3ex\hbox{$\;<$\kern-0.75em\raise-1.1ex\hbox{$\sim\;$}}}
\def\gsim{\raise0.3ex\hbox{$\;>$\kern-0.75em\raise-1.1ex\hbox{$\sim\;$}}}

 \normalsize
\newcommand{\bmat}{\left(\begin{array}}
\newcommand{\emat}{\end{array}\right)}
\newcommand{\be}{\begin{equation}}
\newcommand{\ee}{\end{equation}}
\newcommand{\bea}{\begin{eqnarray}}
\newcommand{\eea}{\end{eqnarray}}

\usepackage{graphics}
\usepackage{url}
\usepackage{amsmath}
\usepackage{hyperref}
\usepackage{psfrag}
\graphicspath{{./Final-figures/}}

\usepackage{epsfig}
\usepackage{psfrag}
\usepackage{xcolor}
\usepackage{color}
\usepackage{amsmath}
\usepackage{relsize}
\title{\boldmath No-scale hybrid inflation with R-symmetry breaking}

\author[a]{Ahmad Moursy,
}

\affiliation[a]{Department of Basic Sciences, Faculty of Computers and Artificial Intelligence,\\  Cairo University, Giza 12613, Egypt}

\emailAdd{a.moursy@fci-cu.edu.eg}

\abstract{In this paper we provide a no-scale supergravity scenario of hybrid inflation with R-symmetry being broken maximally.
We investigate the inflation dynamics in details in both cases of pure F-term hybrid inflation and when adding constant Fayet-Iliopoulos D-terms. The effective inflation potential is asymptotically flat in a region of the parameter space in both cases. 
We explore all regions in the parameter space when discussing the constraints from the observables. 
We point out a connection between inflation, R-symmetry breaking and GUT scales. The moduli backreaction and SUSY breaking effects are investigated in a specific stabilization mechanism. We emphasis that a successful reheating is not affected by R-symmetry breaking, but it has interesting consequences. We study the reheating in flipped GUT model. We argue in favor of $Z_2$ symmetry  associated with flipped GUT models to avoid phenomenologically dangerous operators and allow for decay channels for the inflaton to right-handed neutrinos (sneutrinos).}

\begin{document} 
\maketitle
\flushbottom

%%%%%%%%%%%%%%%%%%%%%%%%%%%%%%%%%%%%%%%%%%%%%%
\section{Introduction}
\label{sec:intro}
%%%%%%%%%%%%%%%%%%%%%%

Since its discovery, the accumulation of the data from the Cosmic
Microwave Background (CMB) over the past years supports the cosmological inflation paradigm. The most recent data  by Planck collaboration \cite{Akrami:2018odb}, confirmed that the spectral index of the scalar fluctuations is $n_s=0.955-0.974$, up to 2 sigma exclusion limits, while the upper bound on the tensor to scalar ratio is  $r<0.08$.  
This may hint at a connection between the
ideas of cosmological inflation and supersymmetric grand unification. 
It turns out that the inflation energy scale is estimated as
\bea
V^{1/4} \sim \left( \dfrac{r}{0.01}\right)^{1/4} \times 10^{16} \,\, {\text GeV},
\eea

One of the key issues in cosmology,
is building a cosmological inflation model that accommodates the current observational constraints and connects to particle physics via reheating phase.
Supergravity offers a promising framework for constructing inflationary models. At such large scale of inflation, supergravity effects should be taken into account. Building supergravity models of inflation is not an easy task. Basically, three problems arise: 
\begin{itemize}
\item The $\eta$-problem: It appears due to the supergravity contributions to the inflaton mass which spoils the slow-roll conditions, $\epsilon\ll 1,\, \eta \ll 1$. More specifically, the expansion of the inflaton scalar potential  yields
$V = V_0 \left( 1+\dfrac{|\phi|^2}{M_P^2}\right)+\cdots$, where $ V_0$ is the inflation scale. Now the Hubble constant squared during inflation is given by $H^2=V_0/ 3 M_P^2$, therefore the inflaton acquires a mass of order Hubble constant which is a generic feature of supergravity models. Hence the slow roll parameter 
$\eta=M_P^2 \left(\dfrac{V''}{V} \right) \approx 1$.

\item Effective single field inflation: On top of its simplicity, it is a sufficient condition for avoiding unacceptably large isocurvature fluctuations. In supergravity models of inflation, the inflaton is not the only scalar field. Even in simple models containing only the inflaton superfield with a complex scalar component, we need only one real degree of freedom to play the role of inflaton and the other being integrated out from the inflation dynamics. Furthermore, in many supergravity models other scalar degrees of freedom appear such as the moduli fields and SUSY breaking superfields in hidden sectors, as well as fields of observable sector containing the Standard model.
Effective single field inflation can be guaranteed if other fields acquire large masses of order Hubble scale and hence frozen during the inflation without affecting the inflation dynamics.

\item Supersymmetry breaking in an approximate flat space: After the end of inflation, the inflaton goes to its true minimum  and
supersymmetry should be broken in an approximate flat space with infinitesimal vacuum energy density $\langle V \rangle \simeq 10^{-120} M_P^4$ according to recent observations that supports a very tiny cosmological constant. This scale is very small compared to the other scales of particle physics. On the other hand connection to low-energy physics imposes a lower bound on the SUSY breaking soft masses $\tilde{m}\gtrsim 10^{-15} M_P$ and on the other hand they are bounded from above by the SUSY breaking scale $M_S=\langle F_I\rangle^{1/2}$. This implies that the supergravity scalar potential $\langle V \rangle \gtrsim 10^{-60} M_P^4$, which contradicts the above tiny value of the cosmological constant. One solution of such problem is to consider SUSY breaking with Minkowski vacuum as in models of no-scale supergravity \cite{Cremmer:1983bf,Ellis:1984bm}. Moreover the SUSY breaking sector has a non trivial backreaction on the inflation potential and may spoil the inflation.

\end{itemize}

The $\eta$-problem in models of supergravity can be solved by defining a shift symmetry on a singlet inflaton \cite{Kawasaki:2000yn,Yamaguchi:2000vm,Brax:2005jv}, and moreover it can be defined on charged inflaton \cite{Heurtier:2015ima,Gonzalo:2016gey}. The K\"ahler potential $K(|X|^2,S+\bar{S})$ is invariant under the shift symmetry $S \to S+ic$, with $c$ is a real constant, while the superpotential takes the form 
\bea
W= X\,f(S),
\eea
where $X$ is a stabilizer field that is introduced to avoid negative quartic terms for large values of the inflaton $S$, and $f(S)$ is a holomorphic function in $S$. The class of models where $f(S)$ is chosen to be a monomial \cite{Kawasaki:2000yn,Yamaguchi:2000vm,Brax:2005jv,Kallosh:2010ug,Heurtier:2015ima,Gonzalo:2016gey} has a common  imprint of having unacceptably large value of tensor to scalar ratio $r \gtrsim 0.1$ which is excluded by Planck recent observations \cite{Akrami:2018odb}. However, models with small tensor to scalar ratio such as the Starobinsky potential of inflation, can be accommodated in the above setup \cite{Kallosh:2013lkr}.

Hybrid inflation models \cite{Linde:1993cn} can connect the inflation physics and particle physics via introducing a GUT gauge symmetry, where the inflaton is coupled to the GUT higgs fields. A supersymmetric model of hybrid inflation, with an exact $U(1)_R$ symmetry (R-symmetry), was introduced in \cite{Dvali:1994ms}. With the following superpotential and minimal K\"ahler potential 
\bea \label{Eq:RS-HI}
W &=& \kappa S(\phi_1 \phi_2 -M^2),\nonumber\\
K &=& |S|^2+ |\phi_1|^2 + |\phi_2|^2,
\eea 
inflation can be realized along the flat direction in which the GUT higgs fields $\phi_1, \phi_2$ are frozen at the origin. The universe is dominated by a constant energy density $V=\kappa^2 M^4$ as long as $|S| > |S_c| =  M$. The Coleman-Weinberg 1-loop correction to the potential provides a slope for the inflaton to slowly roll resulting in small field inflation. The previous superpotential is the most general renormalizable one which is consistent with $U(1)_R$ symmetry. The R-charge assignments are as follows: $R[S]=1$, $R[\phi_1 \phi_2]= 0$ and $R[W]=1$.

In that context, R-symmetry has important advantages. First, it prevents higher degree terms such as $S^2$ and $S^3$ which spoil the small field inflation. Furthermore, the term $\mu_\phi \,\phi_1 \, \phi_2$ is not allowed which spoils the inflation trajectory, $\phi_1= \phi_2=0$, and breaks SUSY. Second it avoids naturally the $\eta$-problem when supergravity corrections are included, since the calculated mass squared of the inflaton from the supergravity potential cancels at the tree level \cite{BasteroGil:2006cm}.
Third, $U(1)_R$ symmetry has many phenomenological advantages in the low energy effective theory \cite{Dvali:1997uq,Lazarides:1998iq,Kyae:2005nv}. It forbids higher dimensional operators that contribute to proton decay as it gives rise to an accidental $U(1)_B$. Moreover its unbroken $Z_2$ subgroup acts as matter parity which prevents couplings that lead to the LSP decay. In addition, R-symmetry may contribute to a solution to the $\mu$-problem as it forbids the MSSM Higgs mixing term $\mu_H H_u H_d$. The latter term can be generated via Giudice-Masiero (GM) mechanism \cite{Giudice:1988yz}.
 Nevertheless, the standard hybrid inflation models with R-symmetry predicts a large spectral index $n_s\sim 0.98$ far from the observation limits, and small tensor to scalar ratio $r\sim 10^{-5}$.

Interestingly, no-scale supergravity offers a natural solution to the $\eta$-problem and can yield an inflation potential with a plateau adequate for slow rolling. Indeed the supergravity scalar potential resembles that in a globally supersymmetric version, where a cancellation occurs between $|W|^2$ and terms of K\"ahler derivatives products in $|D W|^2$. The latter happens due to the noncompact $SU(N,1)/SU(N)\times U(1)$ no-scale symmetry \cite{Cremmer:1983bf,Ellis:1984bm}. Therefore the scalar potential goes like
$$V \sim e^{K/M_{P}^2} \, \sum_\phi \left|\dfrac{\partial W}{\partial \phi}\right|^2.
$$
In some cases \cite{Cecotti:1987sa,Kallosh:2013lkr,Ellis:2013xoa}, the resulting inflationary potential is the Starobinsky potential of inflation \cite{Starobinsky:1980te}, with the predicted inflation observables are in the core of the allowed regions of Planck data. 
The Cecotti model \cite{Cecotti:1987sa} and its modification \cite{Kallosh:2013lkr} depend on the no scale symmetry $SU(2,1)/SU(2)\times U(1)$, with two superfields employed as a stabilizer superfield and inflaton superfield. The model of Ellis-Nanopoulos-Olive (ENO) \cite{Ellis:2013xoa} relies on the no-scale symmetry $SU(2,1)/SU(2)\times U(1)$  with two superfields as well. One superfield corresponds to the inflaton $S$ and the other is identified as a modulus superfield $T$. The superpotential was chosen as the Wess-Zumino model, hence the superpotential and the K\"ahler potential are given by
\bea
W &=& \dfrac{\mu}{2} S^2 - \dfrac{\lambda}{3} S^3,\nonumber \\
K &=& -3 \log \left[ T+\bar{T} - \frac{|S|^2}{3}  \right]
\eea
The modulus field is stabilized at high scale by string theory mechanisms such as the KKLT \cite{Kachru:2003aw}, LVS \cite{Balasubramanian:2005zx} or other mechanisms such as \cite{Ellis:2013nxa,Ellis:1984bs}. Therefore, one ends up with a single field inflation. 

In this class of models R-symmetry needn't be exact to have a successful inflation. As a matter of fact embedding the hybrid inflation, with a superpotential respecting the R-symetry (\ref{Eq:RS-HI}), in an ultraviolet theory containing moduli fields, such as no-scale supergravity, has some difficulties. The stabilized moduli backreact nontrivially on the inflation trajectory and in many cases spoil the inflation \cite{Brax:2006ay,Buchmuller:2013uta}. Furthermore, including the inflaton in the no-scale K\"aher potential requires redefining the fields to have canonical kinetic terms. This implies an effective inflaton potential $\sim \cosh^4 x$ which is too steep for inflation.

R-symmetry breaking in connection to inflation was studied in the literature \cite{Civiletti:2013cra,Khalil:2018iip,Schmitz:2016kyr}. In Ref. \cite{Civiletti:2013cra,Khalil:2018iip}, R-symmetry was allowed to be broken softly by adding a Planck suppressed dimension four operator to the superpotential, while R-symmetry was exact on the tree level. In Ref. \cite{Civiletti:2013cra}, non-canonical k\"ahler potential was considered as well as the superpotential (\ref{Eq:RS-HI}) corrected by a dimension four operator, which results in large field inflation. While in Ref. \cite{Khalil:2018iip} Starobinsky like inflation results due to considering no-scale supergravity with a toy superpotential containing dimension two and dimension four operators.

It is worth mentioning that an exact R-symmetry is a necessary condition for supersymmetry breaking according to Nelson-Seiberg theorem \cite{Nelson:1993nf}. On the other hand breaking SUSY spontaneously in a hidden supergravity sector implies a non-vanishing vev of the superotential $\langle W \rangle \neq 0$, since our universe is associated with an infinitesimally small vacuum
energy. Therefore R-symmetry is broken as $W$ has a non-trivial R-charge.

Our prime aim in this paper is to establish model independent hybrid inflation scenario in no-scale supergravity and waive the R-symmetry constraint applied to the standard hybrid inflation models. We like to stress that both the inflation and the low energy consequences will be consistent with observation when R-symmetry is broken maximally.

R-symmetry can be broken explicitly by adding the terms $\mu S^2$ and $\lambda S^3$ to the superpotential and choose $\mu_\phi$ to vanish.
Alternatively, R-symmetry may be broken spontaneously. The latter may be stemming from a hidden sector containing a superfield $\Psi$, with R-charge $R[\Psi]=-1$, that acquires a non-zero vev. In that case the undesirable term $\mu_\phi \,\phi_1 \, \phi_2 $ can be avoided since the term $\Psi \,\phi_1 \, \phi_2 $ is not allowed as well as any higher order operator $\dfrac{\Psi^n}{M_P^{n-1}} \,\phi_1 \, \phi_2 $, where $M_P$ is the reduced Planck mass. %Nevertheless, a natural alternative for $\mu_\phi$ would be to vanish. 
On the other hand, terms like 
$\nu_1 \, \langle \Psi \rangle S^2$ and $\dfrac{\nu_2}{M_P} \, \langle \Psi \rangle^2 S^3$ are allowed, with $\nu_i$ are dimensionless couplings. This gives rise to an effective superpotential containing the terms $\mu S^2$ and $\lambda S^3$ with the identifications $\mu\equiv \nu_1 \, \langle \Psi \rangle $ and $\lambda \equiv \dfrac{\nu_2}{M_P} \, \langle \Psi \rangle^2 $. The values of $\mu ,\lambda$ depend on the R-symmetry breaking scale $\langle \Psi \rangle\gtrsim 10^{16}$ GeV, $\nu_1 \sim 10^{-1}-10^{-4}$ and $\nu_2 \sim 10^{-1}-10^{-3}$. At such large scale of breaking R-symmetry, the associated R-axion problem does not exist \cite{Nelson:1993nf}.\footnote{The complete details of R-symmetry breaking in the hidden sector is beyond the scope of our paper.}

This paper is organized as follows. In section \ref{sec:model1} we investigate the F-term hybrid inflation in no-scale supergravity. We present a complete analysis of the effective inflation potential and explore all allowed regions of the parameter space versus the Planck limits on the inflation observables. We consider also hybrid inflation with Fayet-Iliopoulos D-term in section \ref{sec:model2} and analyse the effective inflation potential. In section \ref{sec:Moduli-SB} we discuss the SUSY breaking and moduli backreaction on the inflation in both models. Section \ref{sec:reheat} is devoted for discussing the reheating.
 We emphasize on the specific choice of $Z_2$ discrete symmetry as well as a gauge group such as flipped $SU(5)$ to study the reheating phase and some other phenomenological consequences when R-symmetry is not exact. Finally we conclude in section \ref{sec:conclusions}.

%%%%%%%%%%%%%%%%%%%%%%%%%%%%%%%%%%%%%%%%%%%%%%%%%%%%%%%%%%%%%%%%%%%%%%%%%%%%%%%%%%%%%%%%%%%%%%%%
%
\section{No scale F-term Hybrid Infation (FHI)} \label{sec:model1}
We consider the following superpotential which is renormalizable and breaks R-symmetry
\bea\label{Eq:superpotentialF}
W=  \kappa S \left( \phi_1 \, \phi_2-M^2 \right) - \frac{\mu}{2} S^2 + \frac{\lambda}{3} S^3 ,
\eea
Here $S$ is the singlet inflaton superfield and $\phi_1 ,{\phi_2}$ represent conjugate representations of the Higgs  supermultipletes that transform non-trivially under GUT gauge group. The scalar components $\phi_1 ,{\phi_2}$ acquire vevs in the SM neutral direction. The parameter $M$ is the GUT symmetry breaking scale and $\kappa$ is a dimensionless coupling. The parameter $\mu $ determines the scale of the inflation and $\lambda$ is a dimensionless coupling and they are responsible for the R-symmetry breaking in the superpotential. The gauge invariant K\"ahler potential has the no-scale structure
\bea\label{Eq:K1}
K=  -3 \log\left[T+\bar{T} - \frac{|S|^2}{3}  - \frac{|\phi_1|^2}{3} - \frac{|\phi_2|^2}{3} \right],%
\eea
which has the no-scale symmetry $SU(3,1)$. 
The modulus $T$ can be stabilized at high scale \cite{Kachru:2003aw,Balasubramanian:2005zx,Ellis:2013nxa,Ellis:1984bs} with $\langle {\text Re}(T)  \rangle =\tau_0$, $\langle {\text Im}(T)  \rangle =0$.

The total scalar potential is the sum of the F-term and D-term potentials $V=V_F+V_D$. The F-term scalar potential is given by 
\bea
V_F = e^K \left[D_I K^{I\bar{J}} D_{\bar{J}}\overline{W}- 3 |W|^2 \right],
\eea
where $I,J$ run over $T, S , \phi_1,\phi_2$, $K^{I\bar{J}}$ is the inverse of the K\"ahler metric $K_{I\bar{J}}= \dfrac{\partial K}{\partial Z^I \partial Z^{\bar{J}}}$, and $D_I$ is the K\"ahler derivative defined by $D_I=\dfrac{\partial}{\partial Z^I} + \dfrac{\partial K}{\partial Z^I}$. We use the lower case letters $i,j$ to run over the inflation sector fields $S , \phi_1,\phi_2$.
Here and in the rest of the paper we work in the units where the reduced Planck mass $M_P$ is unity. The D-term potential is given by
\bea
V_D = \frac{g^2}{2} {\text Re}f^{-1}_{AB} D^A D^B ,
\eea
%\left(|\phi_1|^2 - |\phi_2|^2 \right)^2
where $f_{AB}$ is the gauge Kinetic function and the indices $A,B$ are corresponding to a representation of the gauge group under which $Z^i$ are charged. The D-term $D^A$ is given by 
\bea
D^A = \frac{\partial K}{\partial Z^i} \, \left(T^A \right)^i_j Z^j,
\eea
with $T^A$ are generators of the GUT gauge group in the appropriate representation. Working in the D-flat direction, the total potential will be given by F-term scalar potential

\be\label{Eq:Ftermpot}
V_F= \dfrac{1}{\Omega^2} \, 
\mathlarger{\mathlarger{\mathlarger{‎‎\sum}}}_{i=1}^{‎3}\left|\frac{\partial W}{\partial Z_i}\right|^2, \hspace{1cm} \Omega =
T+\bar{T} - \frac{|S|^2}{3}  - \frac{|\phi_1|^2}{3} - \frac{|\phi_2|^2}{3}
\ee
It is clear that, the above scalar potential is positive semidefinite. Therefore it has a global minimum which is supersymmetric and Minkowskian, located at 
\bea
\langle S\rangle= 0 \,\,\& \,\,\, \langle \phi_1 \phi_2\rangle= M^2 \,\,\& \,\,\, |\phi_1|=| \phi_2|.
\eea 
In fact $D_i W = W =0$ at the minimum.
Looking at the superotential (\ref{Eq:superpotentialF}) which contains three complex degrees of freedom, one notices that it depends on the combination $\phi_1 \phi_2$. Taking into account that R-symmetry is broken as well as the D-flat direction, $|\phi_1|=| \phi_2|=\rho$, hence one real degree of freedom cancel. It is convenient to parametrize the complex scalar fields in terms of their real components as follows

\bea
 S= \frac{s+i\sigma}{\sqrt{2}}\,\,, \,\,\, \phi_1=\dfrac{\rho}{\sqrt{2}} \, exp\left[i \, \dfrac{\theta+\Sigma}{\sqrt{2} \, M} \right]\,\,, \,\,\, \phi_2=\dfrac{\rho}{\sqrt{2}} \, exp\left[i\, \dfrac{\theta-\Sigma}{\sqrt{2} \, M} \right],
\eea

It is clear that the scalar potential (\ref{Eq:Ftermpot}) depends only on four real degrees of freedom, namely $s, \sigma, \rho, \theta$, while the fifth degree of freedom $\Sigma$ will correspond to the massless goldstone boson which is unphysical and will be eaten by the massless gauge boson to render it massive, hence it will not contribute to the dynamics of inflation. In that representation, the minimum of the potential is located at
\bea
s=\sigma = \theta = 0 \,\,\& \,\,\, \rho = \sqrt{2} M .
\eea
However we rewrite the scalar potential (\ref{Eq:Ftermpot}) in terms of the Cartesian variables $s, \sigma, \alpha, \beta$, with $\alpha +i \beta= \rho e^{i\theta/(2\sqrt{2}M)}$, when discussing the simulation and the mass matrices.
%%%%%%%%%%%%%%%%%%%%%%%%%%%%%%%%%%%%%%%%%%%%%%%%%%%%%%%%%%%%%%%%%%%%%%%%%%%%%%%%%%%%%%%%%
\subsection{Inflation trajectory}
\label{sec:FHI-traj}

Along the inflationary trajectory the potential (\ref{Eq:Ftermpot}) is minimized along the D-flat direction $\phi_1= {\phi_2} = 0$,
and the higgs fields are fixed at the origin during the inflation and we have F-term hybrid inflation (FHI).  
Accordingly the effective inflationary potential will be given by 
\bea\label{eq:infpot}
V_{inf}= \frac{1}{\left(2\tau_0- \frac{|S|^2}{3}\right)^2} \left|\kappa M^2  + \mu S - \lambda S^2\right|^2,
\eea
and the Hubble scale during inflation is 
$
H^2=V_{inf}/3
$.
It is worth mentioning that the effective infationary potential is Starobinsky-like and it is similar  to the one obtained in \cite{Romao:2017uwa}.\footnote{In  \cite{Romao:2017uwa} they assumed the no-scale supergravity realization of Starobinsky like inflation, with adding a Polonyi term to break SUSY along and after inflation. In our scenario, we have Starobinsky like potential due to stabilizing the higgs at the origin with broken SUSY during inflation. After the inflation ends, SUSY is exact at the global minimum. Therefore the phenomenology is different. Moreover, we will give a complete analysis of the potential from the point of view of the phenomenology of our model and will investigate all regimes of the parameter space that are not discussed in \cite{Romao:2017uwa}.}

We turn to discuss the stability of the inflation trajectory resulting in effectively a single field inflation. 
In order to have a canonical kinetic terms for the inflaton $S$ we should have the following field redefinition
\begin{figure}[t!]
	\centering
	\includegraphics[width=0.45\textwidth]{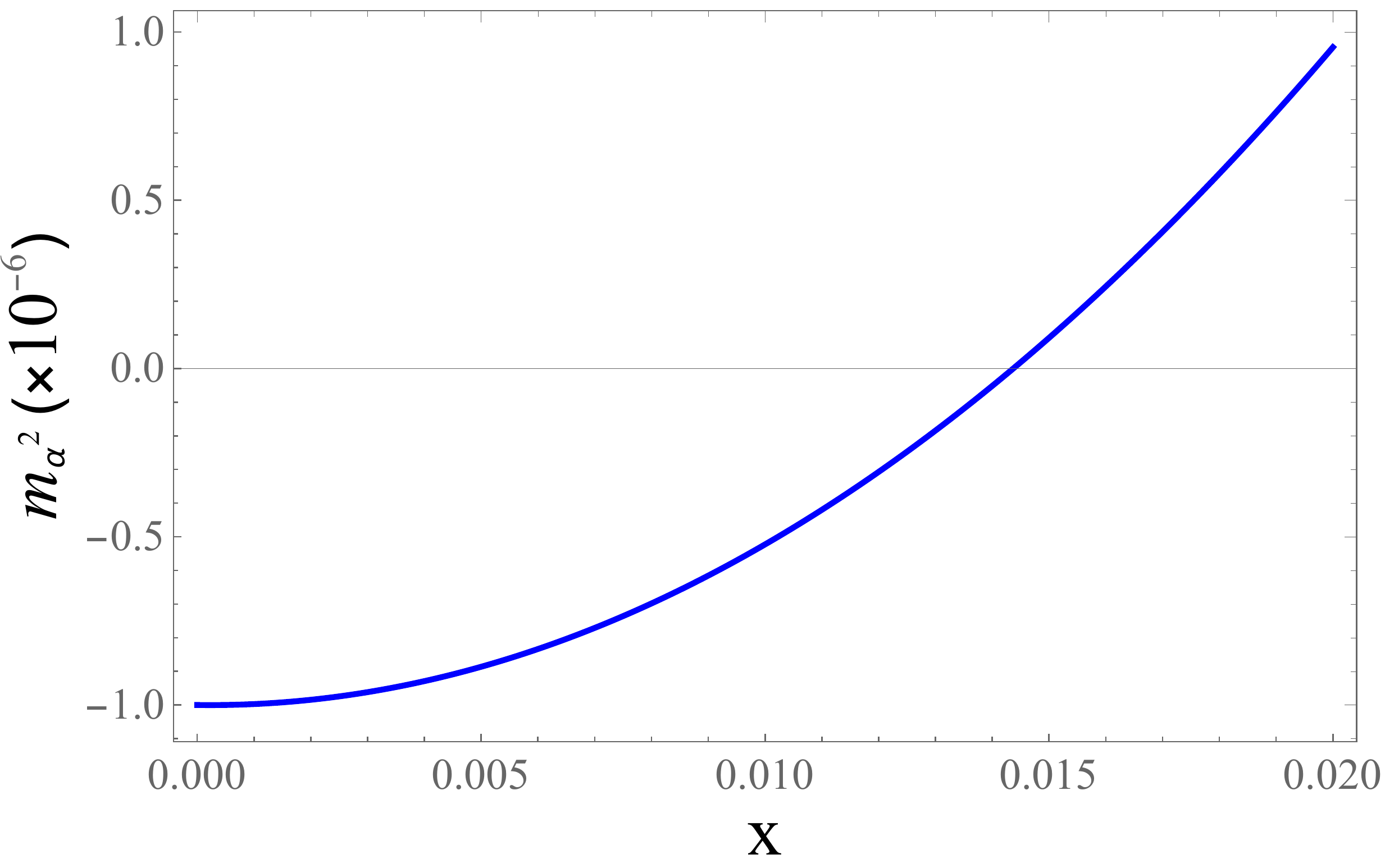}
	\caption{The inflaton dependent mass of the Higgs $\alpha$ with $\lambda=2.2\times 10^{-5}, \mu=3.2 \times 10^{-5} , \kappa=0.1 , M=10^{-2}$. All values are given in the units where $M_P=1$.
	\label{fig:mhiggs2}}
\end{figure}	
\bea\label{Eq:canfield}
S=\sqrt{6\tau_0}\,  \tanh \left( \frac{\chi}{\sqrt{3}}   \right)\, , \, \hspace{1cm} \chi = \frac{ x + i\, y}{\sqrt{2}}\,.
\eea
%
%with 
%$$\chi = \frac{ x + i\, y}{\sqrt{2}},$$
 where $x$ is the slow rolling inflaton. The target space metric during inflation is diagonal in the basis $(x,\alpha,y,\beta)$  and is given by 
 \bea
g_{ij}\Big|_{\text inf}= diag\left[ 1,\dfrac{1}{\tau_0} \cosh ^2\left(\frac{x}{\sqrt{6}}\right) ,1,\frac{1}{\tau_0} \cosh ^2\left(\dfrac{x}{\sqrt{6}}\right) \right]\,.
 \eea
The fields $\alpha , \beta ,y$ are fixed at the origin during the inflation, since the scalar potential is minimized for $\alpha = \beta =y=0$ and their inflaton field dependent masses are larger than the Hubble scale during inflation as follows
\bea
\frac{m_y^2}{H^2} \simeq  4  \, , \, \hspace{1cm}
\frac{m_{\alpha}^2}{H^2}  =    \frac{m^2_{\beta}}{H^2}  \simeq 2\,.
\eea
The above equations have been extracted for large values of the inflaton field $x$. After inflation ends, the fields $\beta , y$ are fixed at zero value. On the other hand, $\alpha$ will be fixed at $\alpha=0$ during inflation as its field dependent mass is positive. As the inflaton rolls down, its value decreases to smaller values until it reaches a critical value $x_c$ at which the field dependent mass $m_{\alpha}^2$ changes to negative and $\alpha=0$ becomes a local maximum as indicated in Fig. \ref{fig:mhiggs2}. This triggers the waterfall phase and $\alpha$ goes to its true minimum $\alpha=\sqrt{2} M$. In particular, for small $x$, to leading order
\bea
m_{\alpha}^2 = \frac{2 \kappa ^2 M^2}{3 c^2} \left( M^2-3 \tau_0\right).
\eea
Therefore, for small values of $x$, $m_{\alpha}^2<0$ whenever $M^2 < 3 \tau_0$.

The critical value of  inflaton $x_c$ which triggers the waterfall, can be computed from 
the inflaton dependent mass squared of $\alpha$ which is  given for small $x$ by 

\bea
m_\alpha^2\simeq \nonumber
 a_0 + a_1 x + a_2 x^2,
\eea
where 
\bea
a_0 &= & -\frac{\kappa ^2 M^2 \left(3 \tau_0 -  M^2\right)}{6 \tau_0^2}  \,, \hspace{1.5cm}
a_1 =     -\frac{\kappa  \mu   \left(3 \tau_0-2 M^2\right)}{6 \tau_0^{3/2}}     \nonumber\\
a_2 &= &     \frac{\left(18 \tau_0^2 \kappa ^2 + 18 \tau_0^2 \kappa  \lambda + 6 \tau_0 \mu ^2 - 3 \tau_0 \kappa ^2 M^2-12 \tau_0 \kappa  \lambda  M^2+4 \kappa ^2 M^4\right)}{36 \tau_0^2}    \nonumber
\eea
Accordingly, the critical value of the inflaton $x_c$ at which the sign of $m_\alpha^2$ flips to a negative sign, is given to leading order in $\mu$ and $\lambda$ by
\bea
x_c \simeq   \sqrt{\frac{1}{\tau_0}} M + \frac{\mu }{\sqrt{4\tau_0}\, \kappa }  - \lambda  \left( \frac{\mu }{\sqrt{4\tau_0} \,\kappa ^2}  +  \frac{M}{\sqrt{4\tau_0}\, \kappa }\right) 
\eea
\begin{figure}[t!]
	\centering
	\includegraphics[width=0.45\textwidth]{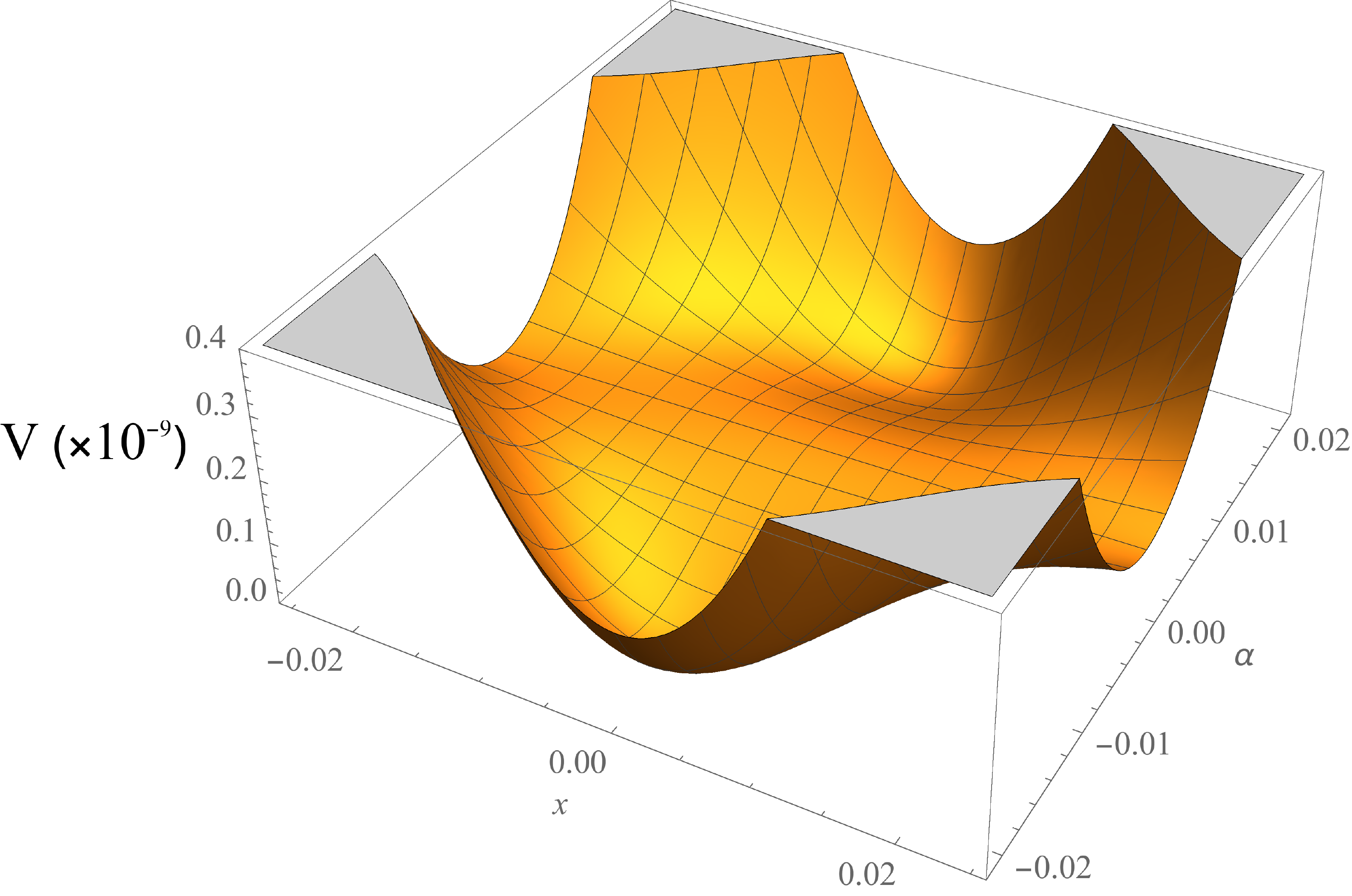}
	\includegraphics[width=0.45\textwidth]{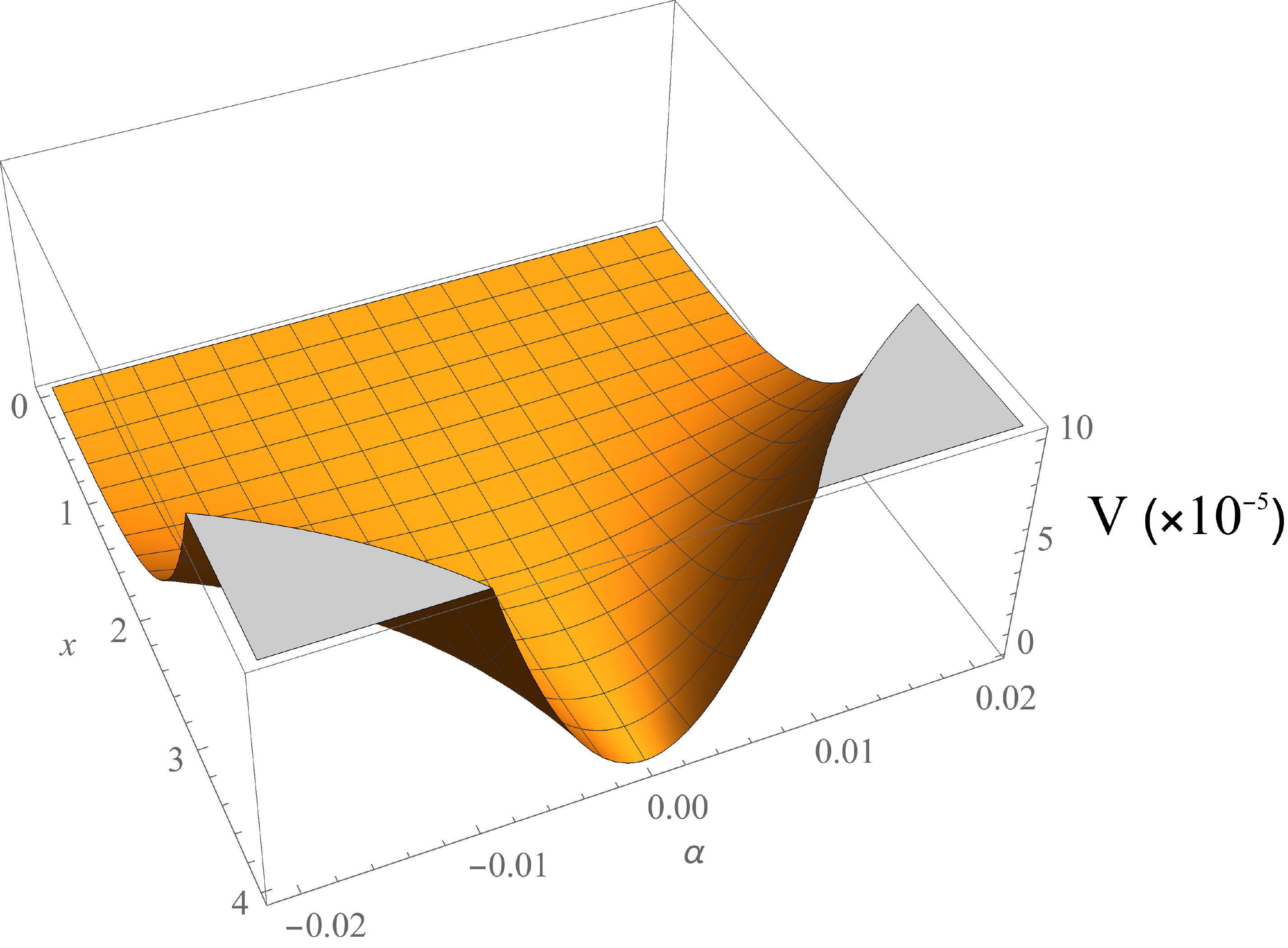}
	\caption{The scalar potential of $x,\alpha$ near the SUSY vacuum (left panel) and  for large values of $x$ (right panel), with $\beta=y=0$ and $\lambda=2.2\times 10^{-5}, \mu=3.2 \times 10^{-5} , \kappa=0.1 , M=10^{-2}$. All values are given in the units where $M_P=1$.
	\label{fig:3Dpot}}
\end{figure}	

In Fig. \ref{fig:3Dpot}, we show the potential of $x,\alpha$. For large values of $x$ the potential is minimum $\alpha$ direction at $\alpha=0$ and the $x$ direction is flat, whereas for small $x$ inflation ends and waterfall happens.

\begin{figure}[h!]
	\centering
	\includegraphics[width=0.49\textwidth]{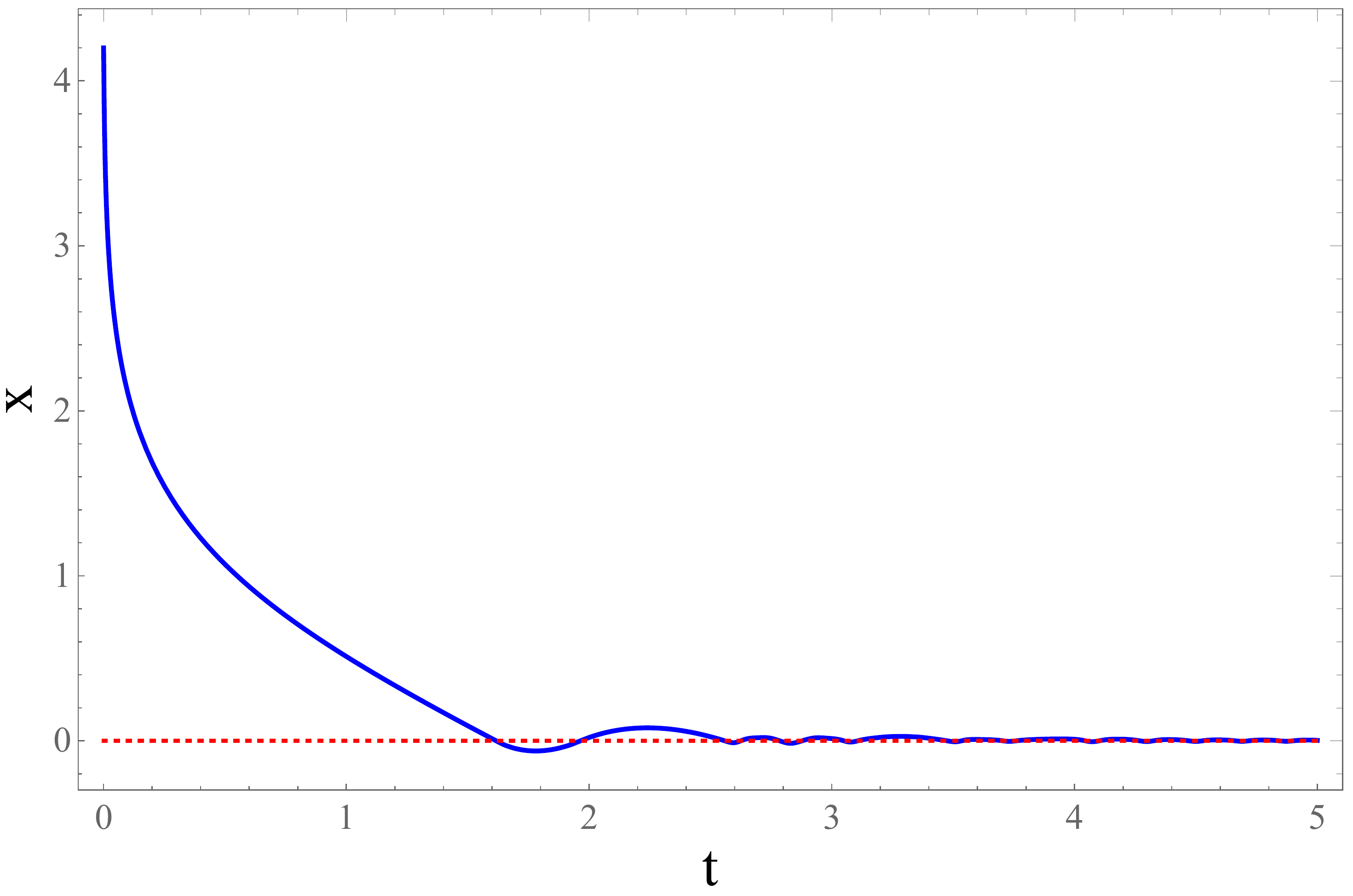}
		\includegraphics[width=0.49\textwidth]{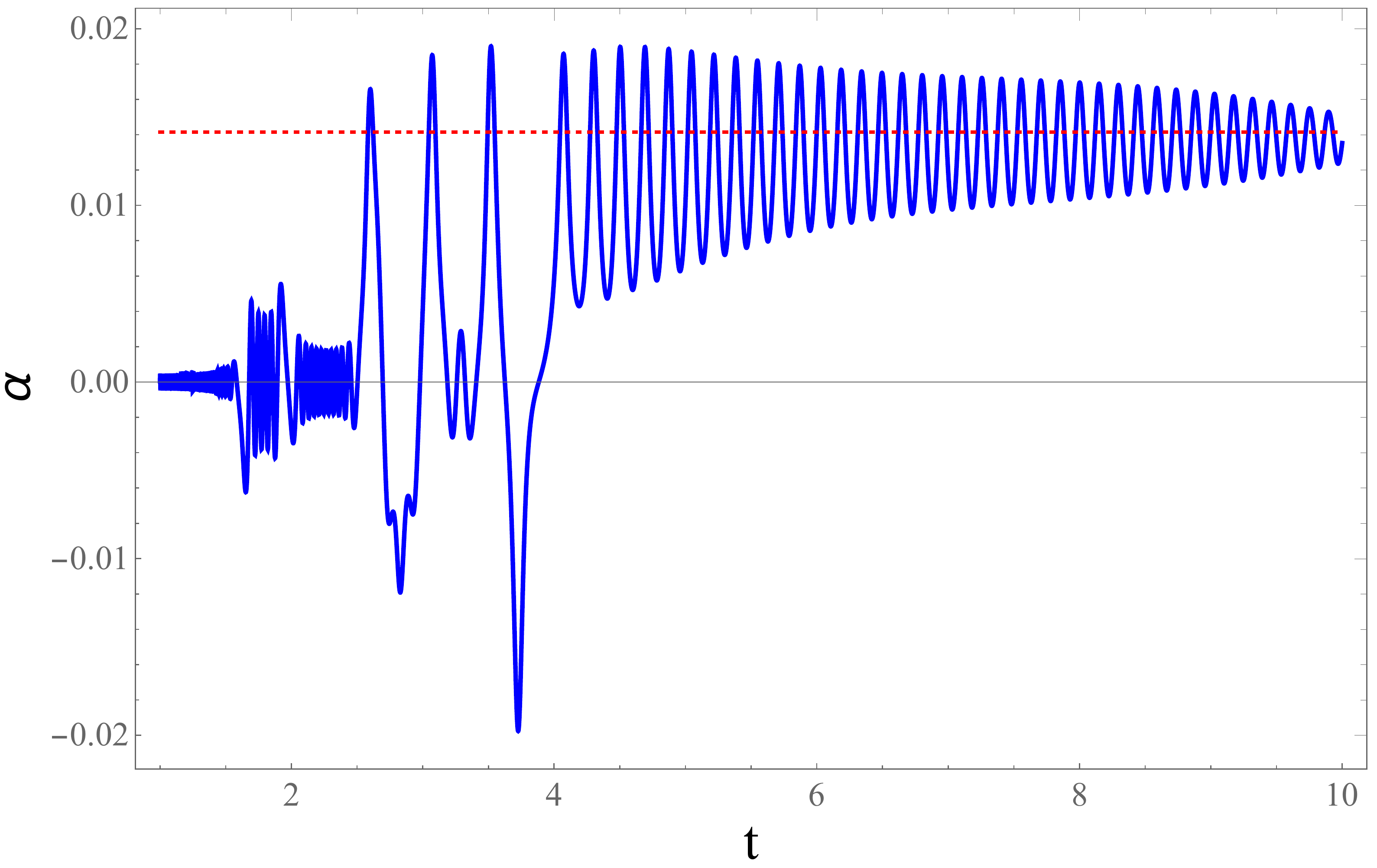}
	\caption{Simulation of the scalar fields: $x(t)$ (left panel) and $\alpha(t)$ (right panel), where $\lambda=2.2\times 10^{-5}, \mu=3.2 \times 10^{-5} , \kappa=0.1 , M=10^{-2}$. The red dotted lines represent the field values at the SUSY minimum. All values are given in the units where $M_P=1$. The time is measured in Hubble units.
	\label{fig:simulation1}}
\end{figure}	

We will simulate the time evolution of the scalar fields by solving the supergravity equations of motion:
\bea\label{Eq:sugraEoM}
\ddot{\Psi}^I \, +  \, + 3H \dot{\Psi}^I + \Gamma^I_{JL} \dot{\Psi}^J \dot{\Psi}^L + K^{I \bar{J}} \dfrac{\partial V}{\partial{\bar{\Psi}}^{\bar{J}}} = 0,\\
H^2=\dfrac{1}{3}\left[  K_{J \bar{L}} \, \dot{\Psi}^J \, {\dot{\bar{\Psi}}}^{\bar{L}}\,  + \, V\left({\Psi}^I\right) \right]\,,
\eea

where $\Gamma^I_{JL}= K^{I \bar{Q}} \partial_J K_{L \bar{Q}}$ are the connection coefficients of the K\"ahler manifold.

Fig. \ref{fig:simulation1} depicts the the simulation  of the inflaton $x$ which slowly rolls and finally reaches the SUSY minimum at $x=0$, while the higgs $\alpha$ is fixed at zero during inflation then acquires a tachyonic mass hence goes to the SUSY minimum at $\alpha=\sqrt{2} M$. On the other hand, $y,\beta$ are fixed at zero during and after inflation. The time has been rescaled by the Hubble constant $H$.
%%%%%%%%%%%%%%%%%%%%%%%%%%%%%%%%%%%%%%%%%%%%%%%%%%%%%%%%%%%%%%%%%%%%%%%%%%%%%%%%%%%%%%%%%%%%%%%%%
\subsection{Inflaton effective potential}
\begin{figure}[t!]
	\centering
	\includegraphics[width=0.5\textwidth]{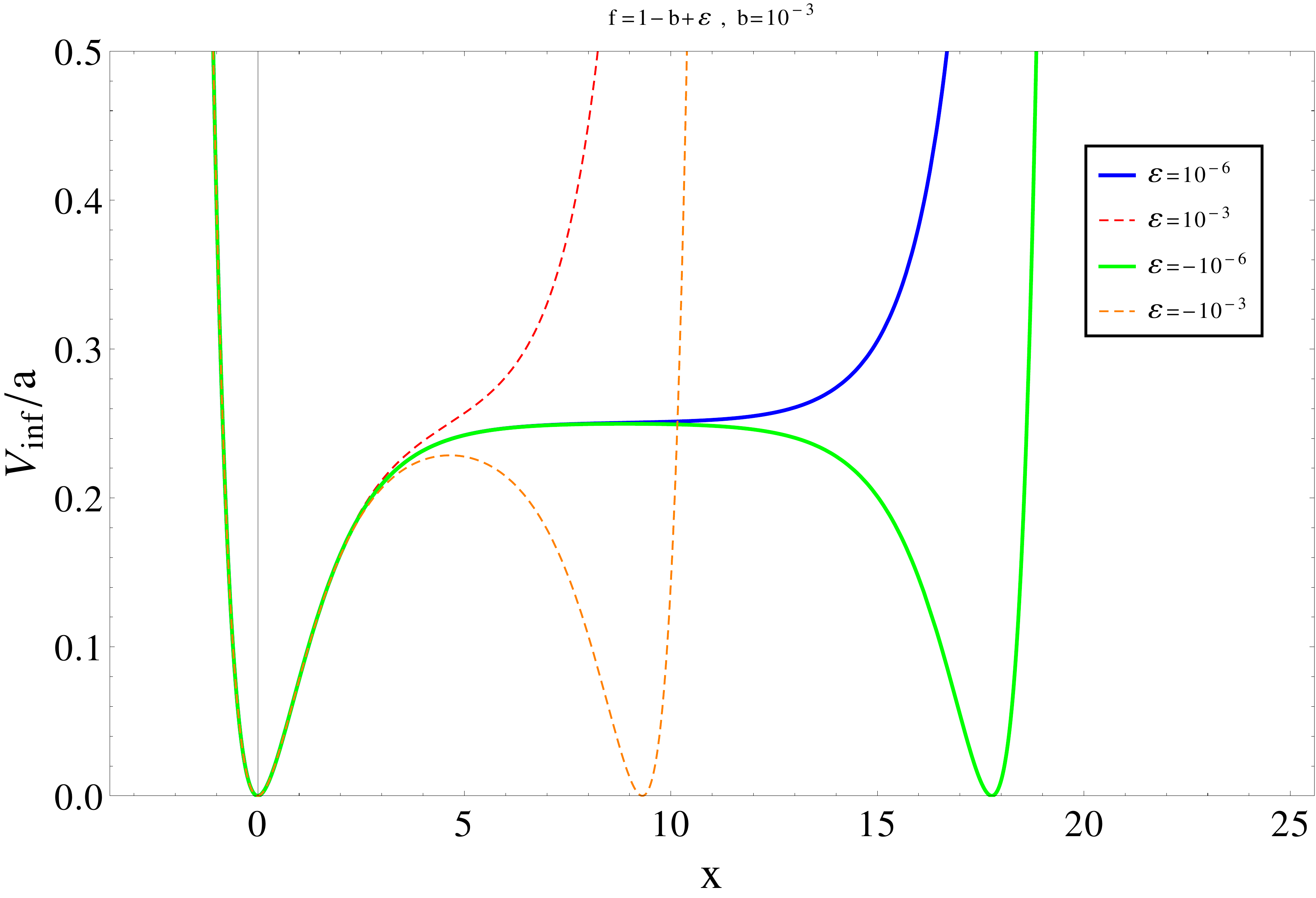}
	\caption{The inflationary potential \ref{Eq:infpot} in case $f=1-b+\varepsilon$, for different non-zero values of $\varepsilon$ and keeping $b$ fixed.
	\label{fig:potepsilon}}
\end{figure}

The potential (\ref{eq:infpot}) is positive semi-definite and its global minimum is $V_{inf}=0$ when 
\bea
S=\frac{1}{2\lambda}\left( \mu\pm \sqrt{\mu^2+4\lambda \kappa M^2 }  \right)
\eea
\begin{figure}[b!]
	\centering
	\includegraphics[width=0.45\textwidth]{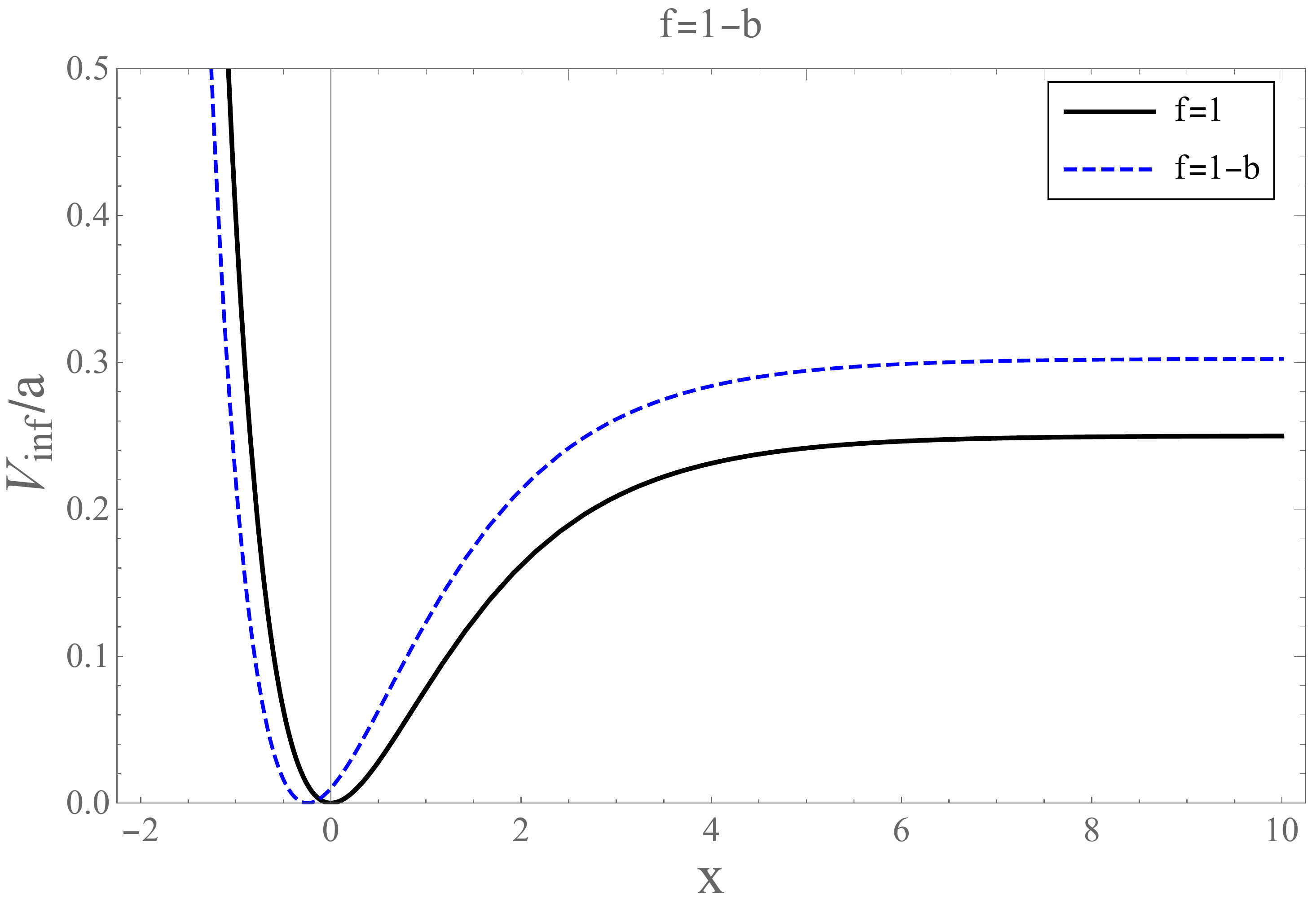}
		\includegraphics[width=0.45\textwidth]{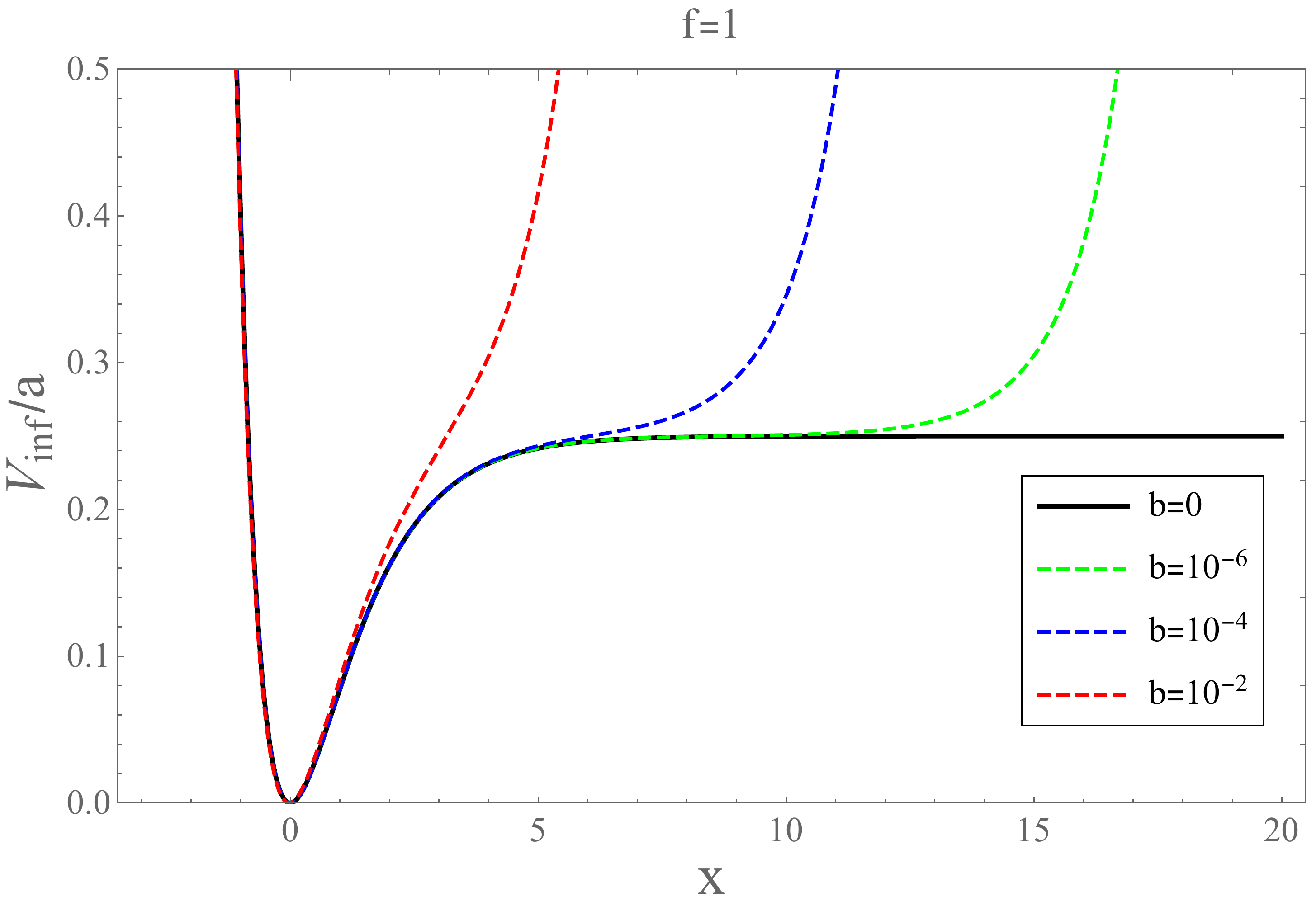}
	\caption{The left panel rerpresents Starobinsky-like potential (\ref{Eq:infpot}) for $b=1-f$. The black curve represents the Starobinsky case when $f=1,\, (b=0)$ and the minimum is located at the origin, while the dashed blue one represents the case when $f$ deviates from 1 by an amount equals $b$ and the minimum is shifted to the left. The potential is flat for large values of $x$.	The right panel  rerpresents Starobinsky-like potential (\ref{Eq:infpot}) with fixing $f=1$ and changing $b=0,10^{-6},10^{-4},10^{-2}$. The shifted minima are not clear due to the scale of the graph.
	\label{fig:potstarlike}}
\end{figure}	
Here, it is clear that for $M=0$, we return to the original ENO model of \cite{Ellis:2013xoa}. 
Using the field redefinition (\ref{Eq:canfield}), the resulting potential will have the form
\bea
V_{\text inf}= a\ \text{sec}^2 \left( \sqrt{\frac{2}{3}} y\right)\ \left| \text{cosh}\left( \frac{x+iy}{\sqrt{6}}\right)  \right|^4\ \left| b + f \tanh \left( \frac{x+iy}{\sqrt{6}}\right) - \tanh^2 \left( \frac{x+iy}{\sqrt{6}}\right) \right|^2 \ 
\eea
where
\bea
a = |3 \lambda |^2 \,, \hspace{0.5cm}
b = \frac{\kappa\, \hat{M}^2}{\lambda } \,, \hspace{0.5cm}
f = \frac{\hat{\mu}}{\lambda }\ ,
\eea
where $\hat{\mu}=\frac{\mu}{\sqrt{6 \,\tau_0}}$ , $\hat{M}=\frac{M}{\sqrt{6 \,\tau_0}}$ are dimensionless quantities.
 Apparently the simple Starobinsky case (ENO) \cite{Ellis:2013xoa}, is recovered for $f=1, b=0$. Expanding the hyperbolic functions, considering that $y$ is frozen at the origin, the effective inflationary potential will be given by

\bea\label{Eq:infpot}
V_{\text inf}=\frac{a}{4} \left( (1+ b) +(b-1) \cosh \left(\sqrt{\frac{2}{3}} x\right)+ f \sinh \left(\sqrt{\frac{2}{3}} x\right)\right)^2
\eea
Clearly, the location of the minimum of the potential at $x_0$ is shifted from the origin once $b\neq 0$:
$
x_0= \sqrt{6} \tanh^{-1} \left(  \dfrac{f\pm \sqrt{f^2+4b}}{2}\right)
$.
The lower sign is chosen such that the minimum is shifted to the left and hence $x_*> x_0$ is guaranteed.
In that respect, we have the important constrain $x_c > x_0$ is satisfied also. The latter allows for the waterfall and hence the fields $\alpha, x$ stabilize to their true minima at $0, \sqrt{2}M$, respectively.
We study two regimes in the parameter space:
\begin{itemize}

{\bf \item Case I}

An interesting case for the inflation potential is the limit when $b\to 1-f$, the potential becomes flat for large values of $x$ with constant hight $\frac{a}{4} (1+b)^2$.\footnote{In \cite{Romao:2017uwa}, they considered only the case where $f=1$ and $b$ is being a perturbation for ENO model. They infer a restriction from inflation on $b \lesssim 10^{-4}$ which gives rise to TeV SUSY breaking scale. It will turn out that $b$ is not restricted by inflation to such tiny values, since the potential in the regime $f = 1-b$ has a plateau for large $x$ and the former constraint is relaxed to be $b\lesssim  0.25$.}
 In this limit the potential will have the following form 
\bea\label{Eq:potstarlike}
%V_{\text inf}\Bigg|_{b \to 1-f}= \frac{a}{4} \left((2-f)-f e^{-\sqrt{\frac{2}{3}} x}\right)^2 
V_{\text inf}\Bigg|_{f \to 1-b}= \frac{a}{4}   \left((b+1) + (b-1)e^{-\sqrt{\frac{2}{3}} x}\right)^2
\eea
Fig. \ref{fig:potstarlike} illustrates the Starobinsky-like potential (\ref{Eq:infpot}) for the case when the parameters $b,f$ are related as $b=1-f$. For the case when $b=0$ ($f=1$), we restore the Starobinsky inflation case. The minimum will be located at 
$
x_0= -\sqrt{6} \tanh^{-1} \left(  b/2\right)
$.

 {\bf \item Case II:}

If $f=1-b+\varepsilon$, the effective potential (\ref{Eq:infpot}) can be written as 
\bea\label{Eq:potII}
V_{eff} &=& \frac{a}{4}   \left[(b+1) + (b-1)e^{-\sqrt{\frac{2}{3}} x}\right]^2
+ \frac{a}{2}  \, \varepsilon \, \sinh \left(\sqrt{\frac{2}{3}} \, x\right) \left[(b+1) + (b-1)e^{-\sqrt{\frac{2}{3}} x}\right]\nonumber\\
&&+\frac{a}{4} \, \varepsilon^2\, \sinh ^2\left(\sqrt{\frac{2}{3}} \,x \right). 
\eea

The first term corresponds to the flat potential (\ref{Eq:potstarlike}) in case I, where $f=1-b$. The second and third terms are corrections due to $\varepsilon$, which are proportional to powers of  $\sinh\left(\sqrt{\frac{2}{3}} \,x \right) $. The latter Spoils the plateau of (\ref{Eq:potstarlike}) as demonstrated in Fig. \ref{fig:potepsilon}.
\end{itemize}

Now we turn to discuss the contributions from radiative corrections to the tree-level scalar potential via one-loop Coleman-Weinberg corrections \cite{Coleman:1973jx}
\begin{equation}
 V_{\text{1-loop}}(S) = \frac{1}{64\pi^2} \text{Str} \left[m(S)^4 \left(\log\left(\frac{m(S)^2}{Q^2}\right) - \frac{3}{2}\right) \right],
\end{equation}
where the supertrace is taken over all superfields with inflaton dependent masses $m(S)$. As advocated above (\ref{sec:FHI-traj}), the stabilized fields during the inflation have $m(S)\sim H$. Since $H^2 \sim a \,M_p^2 \sim 10^{-10} \,M_p^2$, the 1-loop correction $V_{\text{1-loop}} \sim \frac{H^4}{64 \pi^2} \lesssim 10^{-22}$ which is negligible compared to the tree level potential.

%%%%%%%%%%%%%%%%%%%%%%%%%%%%%%%%%%%%%%%%%%%%%%%%%%%%%%%%%%%%%%%%%%%%%%%%%%%%%%%%%%%%%%%%%%%%%%%%%%
\subsection{Inflation Observables}\label{ssec:infobserv}
\begin{figure}[t!]
	\centering
	\includegraphics[width=0.7\textwidth]{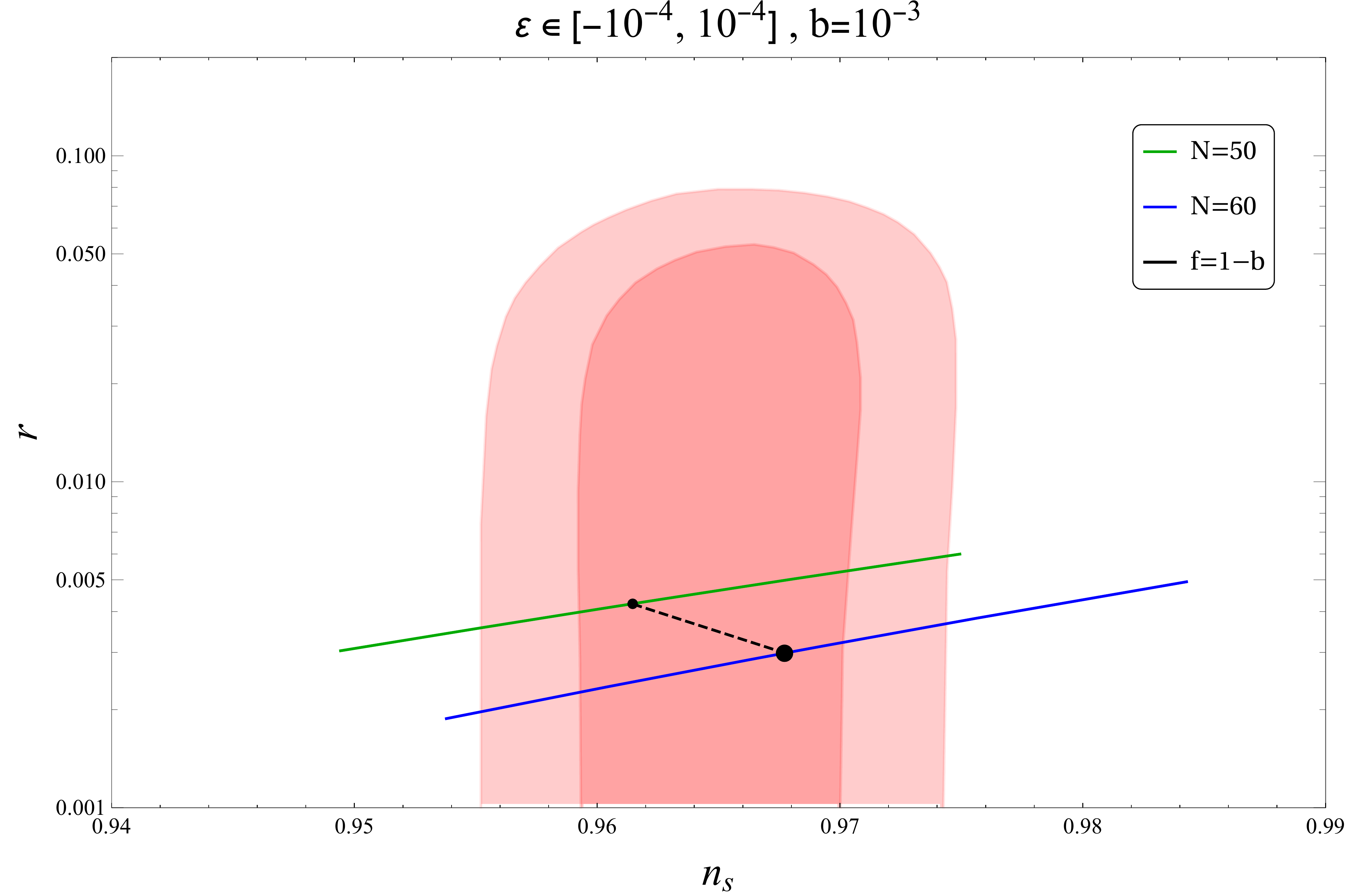}
	\caption{ A logarithmic plot (for vertical axis only) for  $n_s$ and $ r$
 of the inflationary potential Eq. (\ref{Eq:infpot}). Here we fix $b=10^{-3}$ and
scanning over $\varepsilon \in [-10^{-4},10^{-4}]$ with $N=60$ for the blue curve, and $N=50$ for the green curve.
The darker and lighter red regions correspond to the 1 and 2 sigma exclusion limits released by the Planck collaboration (2018)
 (TT+ TE+ EE + LowE + Lensing + BK14)\cite{Akrami:2018odb}.
	\label{fig:Planck1}}
\end{figure}	
Here we will investigate the inflation observables and see the constraints on the different scales $\mu $ and $M$.
We investigate the inflation observables such as the tensor-to-scalar
ratio $r$, scalar tilt $n_s$ and the scalar amplitude $A_s$ (sensitive to the scale of the inflation), and they can be expressed in terms of the slow-roll parameters $\epsilon$ and $\eta$ as follows
\bea
r &=& 16 \epsilon\nonumber \\
n_s&=& 1-6 \epsilon + 2 \eta \nonumber \\
A_s &=& \frac{V}{24 \pi^2 \epsilon},  \nonumber 
\eea
where the above observables are computed at the crossing horizon value of the inflaton field $x_*$. The number of efolding is given by
\bea
N= \int_{x_e}^{x_*} \frac{1}{\sqrt{2 \epsilon}} \, dx,
\eea
where $x_e$ is the value of the inflaton at the end of inflation.
The value of $a$ is fixed by observed value of the scalar amplitude $A_s\simeq 1.95896 \pm 0.10576 \times 10^{-9}$ at 68\% CL  \cite{Akrami:2018odb}.
Now we analyze different regimes in the parameter space that leads to successful inflation.

In the limit $f=1-b$, we analyze the inflation described by effective potential (\ref{Eq:potstarlike}). 
It is clear that the slow roll parameters depend only on $b$. In that case $x_e$ is given by
\bea
x_e= \sqrt{\frac{3}{2}} \log \left(\frac{5 (1-b)}{3 (1+b)}\right).
\eea
This imposes the constrain $ 0 < b < 0.25$, such that $x_e$ is positive. 
The observables $n_s$ and $r$ are independent of $b$, and depend only on $N$ (see Appendix ~\ref{App:starolike}). They have the following form
 
\bea
ns \simeq  %1-\frac{48}{(4 N+2+3 \log (3)-3 \log (5))^2}-\frac{8}{4 N+2+3 \log (3)-3 \log (5)}
 1 - \dfrac{2}{N}-\dfrac{3}{N^2}\,,  \hspace{1cm}
r \simeq  %\frac{192}{(4 \text{NN}+2+3 \log (3)-3 \log (5))^2}
\dfrac{12}{N^2}
\eea

 In that case $b$ has only upper bound is not fixed by $n_s, r $ and $A_s$. 
If the flat regime is perturbed as $f=1-b+\varepsilon$, then corresponding potential (\ref{Eq:potII}) is not asymptotically flat. Hence, the inflation may not succeed.

In Fig. \ref{fig:Planck1} we show a logarithmic plot for $n_s$ and $ r$ prediction
 of the inflationary potential Eq. ~\ref{Eq:infpot}, for both the cases I and II. Changing $b$ doesn't affect the values of $n_s$ and $ r$. In particular, the case of $f=1-b$, represented by the black dashed segment, $n_s$ and $ r$ depends only on $N$ as demonstrated in Appendix ~\ref{App:starolike}. We have changed $f(\varepsilon)$ by changing
 $\varepsilon \in [-10^{-4},10^{-4}]$ for values of  $N=50,60$.

\begin{figure}[h!]
	\centering
	\includegraphics[width=0.6\textwidth]{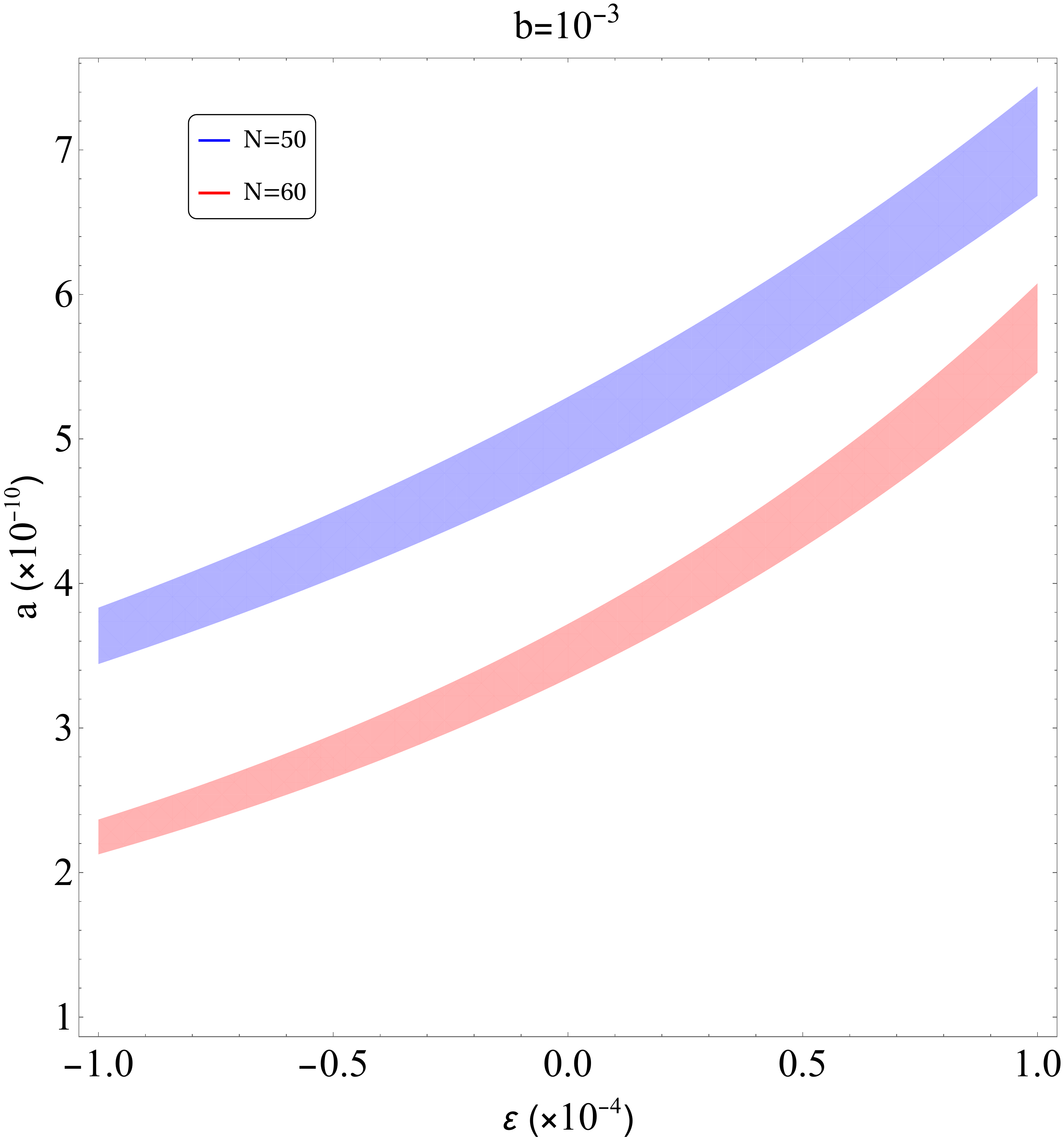}
	\caption{Region plot for the scalar amplitude $A_s$ in the $\varepsilon-a$ plane with fixing $b=10^{-3}$. We let the number of e-foldings $N$ to take two values: $N=50$ for the blue region \& $N=60$ for the red region.
	\label{fig:As}}
\end{figure}

Figure \ref{fig:As} depicts the allowed region by the observed value of scalar amplitude $1.853 \times 10^{-9}\lesssim A_s \lesssim 2.063 \times 10^{-9}$, in the $\varepsilon-a$ plane with fixing $b=10^{-3}$. The range of values of $a \sim 2.2\times 10^{-10}-7.4 \times 10^{-10}$.
The energy  scale of inflation $M_{\text inf}$ (in the flat potential case I) can be estimated in Planck units as
\bea
M_{\text inf} \sim \left( \frac{a}{4} \right)^{1/4} \sqrt{1+b} \sim \, 0.42 \, \left( \dfrac{r}{0.01}\right)^{1/4} \times 10^{-2} 
\sim 3\times 10^{-3}
\eea
Therefore the inflation scale $M_{\text inf}$ is of order GUT scale. For $\kappa\sim 0.1, \, \lambda \sim 2.2\times 10^{-5}, \, \tau_0 \gtrsim 1$ and $b\sim 0.001 - 0.1$, the Grand Unification scale is $M\sim 0.01-0.001$ in Planck units. This provides us with a connection between inflation scale, R-symmetry breaking scale that is encoded in the parameters $\mu$ and $\lambda$, and the GUT scale.
%  

%%%%%%%%%%%%%%%%%%%%%%%%%%%%%%%%%%%%%%%%%%%%%%%%%%%%%%%%%%%%%%%%%%%%%%%%%%%%%%%%%%%%%%%%%%%%%%%%%
\section{No scale Hybrid Infation with constant Fayet-Iliopoulos D-terms (FDHI)} \label{sec:model2}
In this section we add Fayet-Iliopoulos D-term and study the hybrid inflation by considering the same K\"ahler potential (\ref{Eq:K1}) and the following renormalizable superpotential that breaks R-symmetry
\bea\label{Eq:superpotentialD}
W=  \kappa S \, \phi_+ \, \phi_- + \frac{\mu}{2} S^2 - \frac{\lambda}{3} S^3 ,
\eea
Again $S$ is the  singlet inflaton superfield while $\phi_+ ,{\phi_-}$ have opposite charges under $U(1)$ gauge group which is anomalous or non-anomalous \cite{Binetruy:1996xj}. 
The total scalar potential is the sum of the F-term and D-term potentials. Here the D-term $D^A$ is given by 
\bea
D^A = \frac{\partial K}{\partial Z^i} \, \left(T^A \right)^i_j Z^j + \xi^A,
\eea
with $\xi^A$ are Fayet-Iliopoulos D-terms  exist for the $U(1)$ gauge groups. We consider $U(1)$ gauge group and the gauge kinetic function as the Kronecker delta, hence the total potential will be given by 

\be\label{Eq:FDtermpot}
V= \dfrac{1}{\Omega^2} \, 
\mathlarger{\mathlarger{\mathlarger{‎‎\sum}}}_{i=1}^{‎3}\left|\frac{\partial W}{\partial Z_i}\right|^2
+ \frac{g^2}{2}  \left(\dfrac{|\phi_+|^2}{\Omega}- \dfrac{|\phi_-|^2}{\Omega}+\xi \right)^2,
\ee

Again the scalar potential is positive semidefinite. The global minimum is supersymmetric and Minkowskian and is corresponding to $D_i W = W = D^A=0$. It is located at 
\bea
\langle S\rangle= 0 \,\,\& \,\,\, \langle \phi_+ \rangle= 0 \,\,\& \,\,\, \langle | \phi_-|\rangle= \sqrt{\dfrac{6 \tau_0 \xi}{3+\xi}}.
\eea 
We will parametrize the complex scalar fields in terms of their real components as follows

\bea
 S= \frac{s+i\sigma}{\sqrt{2}}\,\,, \,\,\, \phi_+=\dfrac{\alpha_1 + i \beta_1}{\sqrt{2}} \,\,, \,\,\, \phi_-=\dfrac{\alpha_2 + i \beta_2}{\sqrt{2}} ,
\eea

One can write $\phi_-$ in polar representation as $\phi_-=\frac{\rho+\rho_0}{\sqrt{2}}\, e^{\theta/\sqrt{2} \rho_0 }$ with $\rho_0$ being the vev, hence $\theta$ will correspond to the massless goldstone boson and the dynamics will depend on five real degrees of freedom \cite{Heurtier:2015ima}. However will work on the basis $s, \sigma, \alpha_1,\alpha_2, \beta_1,\beta_2$ and $\beta_2$ is mainly the goldstone boson, hence the minimum of the potential is located at
\bea
s=\sigma = \alpha_1=\beta_2=\beta_1 = 0 \,\,\& \,\,\,  \alpha_2= \sqrt{\dfrac{12 \tau_0 \xi}{3+\xi}}.
\eea
%

%%%%%%%%%%%%%%%%%%%%%%%%%%%%%%%%%%%%%%%%%%%%%%%%%%%%%%%%%%%%%%%%%%%%%%%%%%%%%%%%%%%%%%%%%%%%%%%%%
\subsection{Inflation trajectory and effective potential}
\label{ssec:Dtermtraj}
The scalar potential is minimized in the direction $\phi_+=\phi_-=0$ and the effective inflation potential is given by
\bea\label{Eq:Dinfpot}
V_{inf}= \frac{g^2 \xi^2}{2} + \frac{1}{\left(2\tau_0- \frac{|S|^2}{3}\right)^2} \left| \mu S - \lambda S^2\right|^2.
\eea
Clearly, the above potential is the same as the ENO model \cite{Ellis:2013xoa} but shifted by the energy density $\frac{g^2 \xi^2}{2}$.
We use the same field redefinition (\ref{Eq:canfield}), then the target space metric during inflation is found to be diagonal in the basis $(x,\alpha_1,y,\beta_1,\alpha_2,\beta_2)$  and is given by 
 \bea
g_{ij}\Big|_{\text inf}= diag\left[ 1,\dfrac{1}{2 \tau_0} \cosh ^2\left(\frac{x}{\sqrt{6}}\right) ,1,\frac{1}{2 \tau_0} \cosh ^2\left(\dfrac{x}{\sqrt{6}}\right) ,  \frac{1}{2 \tau_0} \cosh ^2\left(\dfrac{x}{\sqrt{6}}\right) ,\frac{1}{2 \tau_0} \cosh ^2\left(\dfrac{x}{\sqrt{6}}\right)   \right]. \nonumber
 \eea

Similarly the fields $\alpha_1 , \beta _1, \alpha _2, \beta  _2,y$ are fixed at the origin during the inflation, since the scalar potential is minimized for $\alpha_1 = \beta_1= \alpha_2= \beta_2 = y= 0$. As a matter of fact, minimizing the potential in the direction of $\alpha_2$ gives two solutions, namely $\alpha_2=0$ and the other solution for $x\gg 1$ is given by $\alpha_2^2 = -\dfrac{24  \tau_0 (\lambda -\hat{\mu} )^2}{\kappa ^2}$. The latter gives complex value for  $\alpha_2$ and the only allowed minimum during inflation is  $\alpha_2=0$. The field dependent masses are larger than the Hubble scale during inflation as follows
\bea
\frac{m_y^2}{H^2} \simeq  4 , \hspace{1cm}
\frac{m_{\alpha_1}^2}{H^2} =    \frac{m^2_{\beta_1}}{H^2}  = \frac{m_{\alpha_2}^2}{H^2} =   \frac{m^2_{\beta_2}}{H^2}  \simeq 2.
\eea
The above equations have been extracted for large values of the inflaton field $x$.
In fact the field dependent squared mass matrices of $(\alpha_1,\alpha_2)$ and $(\beta_1,\beta_2)$ have mixing terms which are very small in the large limit of the inflaton field.
 The fields $\alpha_1,\beta_1,\beta_2 , y$ will be fixed at zero during and after inflation. On the other hand $\alpha_2$ will be fixed at $\alpha_2=0$ during inflation with positive field dependent mass squared. As the inflaton rolling down, its value decreases to smaller values until it reaches a critical value $x_c$ at which the field dependent mass $m_{\alpha_2}^2$ changes to negative and $\alpha_2=0$ becomes a local maximum. This triggers the waterfall phase and $\alpha_2$ goes to its true minimum. In particular, for small $x$, to leading order 
$
m_{\alpha_2}^2 = -g^2 \xi 
$.
 To find the critical value of the inflaton $x_c$ which triggers the waterfall, we expand the masses for small $x$, hence 

\bea
m_{\alpha_2}^2\simeq 
 x^2 \left(\frac{\kappa ^2}{2}+\hat{\mu} ^2\right)-g^2 \xi \,.
\eea
Accordingly, the critical value $x_c$ at which the sign of $m_\alpha^2$ flips to negative sign, is given to leading order in $\xi$ by
\bea
x_c \simeq   \frac{ g \sqrt{2\xi }}{\sqrt{\kappa ^2+2 \hat{\mu} ^2}}\,.
\eea
%
%Fig. \ref{fig:Dtermmhiggs2} shows the flipping if the sign of the Higgs mass squared after $x_c$.

%
\begin{figure}[t!]
	\centering
	\includegraphics[width=0.6\textwidth]{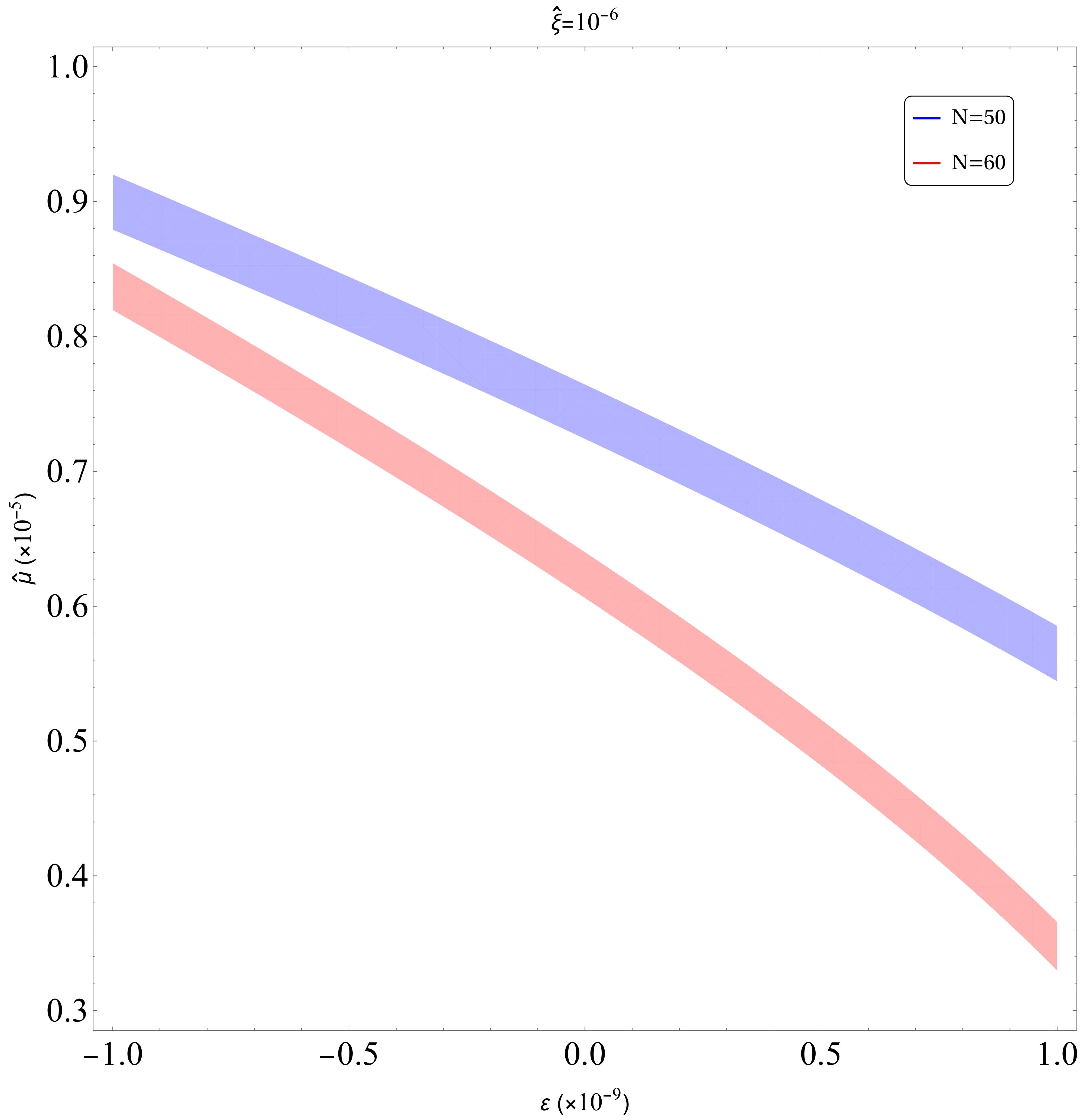}
	\caption{Region plot for the scalar amplitude $A_s$ in the $\varepsilon-a$ plane with fixing $\hat{\xi}=10^{-6}$. We let the number of e-foldings $N$ to take two values: $N=50$ for the blue region and $N=60$ for the red region.
	\label{fig:DtermAs}}
\end{figure}	

The effective inflationary potential has a plateau for $\lambda = \hat{\mu}$ and is given by
\bea\label{Eq:Dterm-vinf1}
V_{\text inf}= \hat{\xi}^2 + \frac{9 \hat{\mu} ^2}{4}   \left(1-e^{-\sqrt{\frac{2}{3}} x}\right)^2,
\eea
which is the Starobinsky potential shifted by $\hat{\xi}^2=\frac{g^2 \xi ^2}{2}$. 
On the other hand, perturbing the plateau with $\lambda = \hat{\mu}+\varepsilon$, the potential is given by
\bea\label{Eq:Dterm-vinf2}
V_{\text eff}&=& \hat{\xi}^2 + \frac{9 \hat{\mu} ^2}{4}   \left(1-e^{-\sqrt{\frac{2}{3}} x}\right)^2 
-18 \,\varepsilon \,  \mu \, e^{\sqrt{\frac{2}{3}} x} \, \sinh^3\left({\sqrt{\frac{2}{3}} x}\right)
+ 9\, \varepsilon ^2 \, e^{2\sqrt{\frac{2}{3}} x} \sinh^4\left({\sqrt{\frac{2}{3}} x}\right) \nonumber\\
.
\eea
Therefore the plateau is spoiled by the last two terms which are very steep. Here we stress that the infation potentials~(\ref{Eq:Dterm-vinf1},\ref{Eq:Dterm-vinf2}) are only valid for $x>x_c$, otherwise the potentials are minimized at $x=0$ with cosmological constant of order $\hat{\xi}^2$ which is inconsistent with the observation of infinitesimally small cosmological constant.

Now we turn to discuss the observables. The value of $x$ at the end of inflation is
\bea
x_e=\sqrt{\frac{3}{2}} \log \left(\frac{15 \hat{\mu} }{2 \sqrt{36 \hat{\mu} ^2+15 \hat{\xi} ^2}-3 \hat{\mu}}\right).\nonumber
\eea
In order to have $x_e > 0$, the constraint $\hat{\xi}<\sqrt{3} \hat{\mu}$ should be satisfied. The shift by  $\hat{\xi}^2$ doesn't alter the predictions of the ENO model \cite{Ellis:2013xoa}, whenever $\hat{\xi}<\sqrt{3} \hat{\mu}$. The scale of inflation is determined by 
observed value of the scalar amplitude $A_s\simeq 1.95896 \pm 0.10576 \times 10^{-9}$ \cite{Akrami:2018odb}, and is given by
\bea
M_{inf}=  \left(\hat{\xi} ^2+ \frac{9}{4} \hat{\mu} ^2 \right)^{1/4}. 
\eea
Since the predicted $r\sim 10^{-3}$, $M_{inf}$ is of order GUT scale.
Figure \ref{fig:DtermAs} displays the allowed region by the observed value of scalar amplitude, in the $\varepsilon-\hat{\mu}$ plane with fixing $\hat{\xi}=10^{-6}$. The range of values of $\hat{\mu} \sim 3.2\times 10^{-5}-9.2 \times 10^{-5}$.

%%%%%%%%%%%%%%%%%%%%%%%%%%%%%%%%%%%%%%%%%%%%%%%%%%%%%%%%%%%%%%%%%%%%%%%%%%%%%%%%%%%%%%%%%%%%%%%%%
\section{Moduli backreaction and SUSY breaking}
\label{sec:Moduli-SB}
An essential component of the no-scale inflationary models is the modulus field $T$. It turns out that the stabilization mechanism of the modulus field can affect the inflation trajectory \cite{Brax:2006ay,Brax:2006yq,Davis:2008fv,Linde:2011ja,Buchmuller:2013uta,Buchmuller:2014vda,Buchmuller:2014pla,Davis:2008fv,
Buchmuller:2015oma,Wieck:2014xxa,Dudas:2015lga}. 
In this section we study the effect of the modulus stabilization and the expected backreaction on the inflationary trajectory. We will focus on the mechanism proposed in \cite{Ellis:2013nxa, Ellis:1984bs} which provides a strong stabilizing terms in the K\"ahler potential
as follows
\bea\label{Eq:K2}
K=  -3 \log\left[T+\bar{T} - \frac{|S|^2}{3}  - \frac{|\phi_1|^2}{3} - \frac{|\phi_2|^2}{3} + \dfrac{\left( T+\bar{T} -2 \tau_0  \right)^4 + \left( T - \bar{T}   \right)^4}{\Lambda^2}\right],%
\eea
where the scale $\Lambda \ll 1$ such that the modulus acquires large mass and stabilizes during the inflation. The above K\"ahler potential preserves the no-scale structure with stabilizing the modulus at $\tau_0$, where $\tau_0$ represents the minimum of the modulus in absence of inflation sector. Including the inflation sector, the large positive energy density during inflation shifts the modulus minimum of $T$ by $\delta T$. The effective scalar potential is then given, to leading order in $ \delta T$, $\delta \bar{T} $, in terms of the total supergravity scalar potential $V$ by
\bea
V_{\text eff}= V(\tau_0) + \left(\delta T \,\dfrac{\partial V}{\partial T}\Big|_{\tau_0} + \delta T ^2\, \dfrac{\partial^2 V}{\partial T^2}\Big|_{\tau_0} +c.c. \right) +  \delta \bar{T} \,  \delta T \,\dfrac{\partial^2 V}{\delta \bar{T} \,\partial T}\Big|_{\tau_0} 
+{\cal O}(\delta T ^3)
\eea
The displacement $\delta T $ is obtained by imposing the minimization condition: $\partial_T V |_{\tau_0+\delta T} = 0$ or equivalently $\partial_{\delta T} V_{\text eff} = \partial_{\delta \bar{T}} V_{\text eff}  = 0$.

It turns out that the exact no-scale symmetry will preserve the inflation potential from dangerous terms such as the soft mass term and the term proportional to $-3 |W|^2$ \cite{Buchmuller:2015oma,Dudas:2015lga}. However, the stabilizing term in the K\"ahler (\ref{Eq:K2})  doesn't have an origin from UV theory such as string theory. If instead we used mechanisms of moduli stabilization in string theory such as KKLT or LVS models, the no-scale structure is broken by the non-perturbative terms in the moduli superpotential and the backreaction of the moduli results in dangerous terms arising from $|W|^2$ in the scalar potential, which spoils the plateau \cite{Dudas:2015lga}.

%%%%%%%%%%%%%%%%%%%%%%%%%%%%%%%%%%%%%%%%%%%%%%%%%%%%%%%%%%%%%%%%%%%%%%%%%%%%%%%%%%%%%%%%%%%%%%%%%
\subsection{Backreaction on no-scale FHI}
\label{ssec:moduli-FHI}
We add a constant $W_0$ to the superpotential (\ref{Eq:superpotentialF}) and use the K\"ahler potential (\ref{Eq:K2}).
At the global minimum, SUSY is broken via the F-term only in the directions of $T$ and $\phi_1, \phi_2$ with zero cosmological constant. The gravitino mass and the modulus mass are given by 
\bea
m_{3/2}=\frac{W_0}{\left(2 \tau_0-\frac{2 M^2}{3}\right)^{3/2}} \,, \hspace{1cm}  
 m_T^2=\frac{864 \, \tau_0^2 \, W_0^2}{\Lambda^2 \left(M^2-3 \tau_0\right)^2} \simeq  
 \frac{768 \, m_{3/2}^2 \,  \tau_0^3  }{ \Lambda^2 }   \,.
\eea

On the other hand during the inflation SUSY is broken via D-term and via F-term in the direction of $ T $ and $ S $. 
The waterfall fields and $y$ are still fixed at the origin during the inflation. Hence, the effective potential is given by
\bea\label{Eq:VeffF}
V_{\text eff}= V_0 \left[1- \dfrac{8 V_0}{3  m_T^2} +{\cal O}\left(\dfrac{H^3}{m_T^3} \right) \right]\,, \hspace{1cm}
V_0 = \frac{a}{4}   \left((b+1) + (b-1)e^{-\sqrt{\frac{2}{3}} x}\right)^2 \,,
\eea
As expected, the modulus backreaction on the inflation potential results in corrections suppressed by powers of the large  modulus mass which doesn't affect the plateau.

%%%%%%%%%%%%%%%%%%%%%%%%%%%%%%%%%%%%%%%%%%%%%%%%%%%%%%%%%%%%%%%%%%%%%%%%%%%%%%%%%%%%%%%%%%%%%%%%%
\subsection{Backreaction on no-scale FDHI}
\label{ssec:moduli-DHI}
Similarly we add a constant $W_0$ to the superpotential (\ref{Eq:superpotentialD}) and use the K\"ahler potential (\ref{Eq:K2}).
At the global minimum, SUSY is broken via the F-term only in the direction of $T$ and $\phi_-$ with zero cosmological constant. The gravitino mass and the modulus mass are given by 
\bea
m_{3/2}= %\dfrac{W_0}{\left(2 \tau_0-\frac{2 \xi \tau_0}{\xi +3}\right)^{3/2}}
 \left( \dfrac{\xi +3}{6 \tau_0} \right)^{3/2} W_0 \,, \hspace{1cm}  m_T^2=\frac{4 (\xi +3)^2 \left(g^2 \Lambda^2 \xi ^2+72\, W_0^2\right)}{27 \Lambda^2}  \,.
\eea

On the other hand during the inflation SUSY is broken via D-term and via F-term in the direction of $T$. The effective inflation potential is given by 
\bea\label{Eq:VeffD}
V_{\text eff}= \hat{\xi}^2 + V_0 \left[1- \dfrac{8 V_0}{3  m_T^2} +{\cal O}\left(\dfrac{H^3}{m_T^3} \right) \right]\,, \hspace{1cm}
V_0 = \frac{9 \hat{\mu} ^2}{4}   \left(1-e^{-\sqrt{\frac{2}{3}} x}\right)^2 \,,
\eea
where the waterfall fields and $y$ are still fixed at the origin during the inflation. 
Again, the modulus backreaction on the inflaton potential is negligible and the plateau is not affected.

%%%%%%%%%%%%%%%%%%%%%%%%%%%%%%%%%%%%%%%%%%%%%%%%%%%%%%%%%%%%%%%%%%%%%%%%%%%%%%%%%%%%%%%%%%%%%%%%%
\section{Reheating}
\label{sec:reheat}
In this section we study the reheating after inflation which is one of the interesting consequences of the FHI model. The mass matrix of the inflaton and the higgs  (\ref{Eq:FHImass}) is not diagonal. The mixing between the inflaton and the higgs fields is proportional to $\mu$. The later is stemming from R-symmetry breaking term in the superpotential (\ref{Eq:superpotentialF}). This will have important impacts on the reheating scenario. Indeed that provides additional motivation for the inflation scenario ~\ref{sec:model1} where a natural coupling of the inflaton (via the mixing with the GUT higgs) to the the MSSM sector can arise and may contribute to the reheating stage.
The mixing angle is given, in terms of the mass matrix (\ref{Eq:FHImass}) entries,  by
\bea
\tan (2\theta)= \dfrac{2 {\cal M}_{12}}{{\cal M}_{22} - {\cal M}_{11}}
\eea
The physical states (we use the canonical inflaton $x$) and the physical masses are given as follows
\bea
x'=x \,\cos \theta \, +\, \alpha \, \sin \theta \,\,, \hspace{1cm}
\alpha' = -x \,\sin\theta \, + \, \alpha \,\cos \theta \nonumber%\\
\eea  
%{\nu}_H^c

The reheating is dependent on the choice of the gauge symmetry group.
We will consider the flipped GUT (FGUT) gauge group $SU(5)\times U(1)_X$ (or flipped $SU(5)$) which has many appealing features as well as the advantage of being free from the monopoles \cite{tHooft:1974kcl}. Inflation and reheating in the context of FGUT scenarios was considered in \cite{Gonzalo:2016gey,Kyae:2005nv,Ellis:2017jcp,Ellis:2018moe,Ellis:2019opr}. In \cite{Gonzalo:2016gey} the inflaton was right-handed sneutrino that is charged under FGUT gauge group which allows for natural decay channels to the MSSM particles, while in \cite{Ellis:2017jcp,Ellis:2018moe,Ellis:2019opr} the inflaton was a singlet. In the latter scenario, it was pointed out a connection between inflation, neutrino masses and reheating via the couplings between the inflaton, the right handed neutrino and the higgs. On the other hand, FGUT hybrid inflation with singlet inflaton was explored in \cite{Kyae:2005nv}.

The field representations of the flipped $SU(5)$ group are  listed in Table \ref{tab:U1-Z2} as well as the respective $U(1)_X$ charges.
The $Q_X$ charges are assigned such that the SM hypercharge is obtained as \cite{Gonzalo:2016gey}
\bea
  Y = \frac{1}{5} \left(Q_X - Q_{Y'}\right),
 \label{hypercharge}
\eea
where $Q_{Y'}$ is the charge associated with the first abelian factor of the broken $U(1)_{Y'} \times U(1)_X$, subalgebra of $SU(5)\times U(1)_X$. $Y'$ is the diagonal generator of $SU(5)$ and $Q_{Y'} =\dfrac{1}{6}{\text diag} \left( -2,-2,-2,3,3 \right)$.

 In this regard the particle content is accommodated in the flipped $SU(5)$ representations as follows \cite{Kyae:2005nv,Ellis:2014xda,Gonzalo:2016gey}:

\begin{itemize}
\item The standard model (SM) matter content is contained in the representations $\mathbf{10}_F$, $\mathbf{\bar 5}_F$ and $\mathbf{1}_F$ as follows
\bea
\mathbf{10}_{F}(1)= \{ Q, d^c, \nu^c \}\, , \,\,\,\, \mathbf{\overline{5}}_{F}(-3)=\begin{pmatrix}
u^c \\
L
\end{pmatrix}  \, , \,\,\,\, 
\mathbf{1}_{F}(5)=e^c. \nonumber
\eea

\item The SM Brout-Englert-Higgs bosons responsible for the electroweak symmetry breaking, are contained in $\mathbf{\bar 5}_{H_u}$ and $\mathbf{ 5}_{H_d}$.
\bea
 \mathbf{\bar{5}}_{H_u}(2)=\begin{pmatrix}
D^c \\
H_u
\end{pmatrix}  \, , \,\,\,\, 
\mathbf{{5}}_{H_d}(-2)=\begin{pmatrix}
\bar{D}^c \\
H_d
\end{pmatrix}. \nonumber
\eea

\item The representations $\mathbf{10}_{H}$ and $\mathbf{\overline{10}}_{H}$ trigger the breaking of the flipped $SU(5)$ to the MSSM gauge group by acquiring vevs in the SM neutral direction ${\nu}^c_H, \bar{\nu}^c_H$ that are identified with the field $\alpha$:
\bea
\mathbf{10}_{H}(1)= \{ Q_H, d_H^c, \nu_H^c \}\, , \,\,\,\, \mathbf{\overline{10}}_{H}(-1)= \{ \bar{Q}_H, \bar{d}_H^c, \bar{\nu}_H^c \} \,.
 \nonumber
\eea

\item The inflaton $S$ is assigned to a singlet $\mathbf{1}_{S}$. 

\end{itemize}

\begin{table}[h]
 \centering
 \begin{tabular}{c | c c c c c c c c c}
  \hline \hline
   &$\mathbf{\bar 5}_{H_u}$ & $\mathbf{ 5}_{H_d}$ & $\mathbf{\bar 5}_{F}$ & $\mathbf{10}_{H}$ & $\mathbf{\overline{10}}_{H}$ & $\mathbf{10}_{F}$  & $\mathbf{1}_{F}$ & $\mathbf{1}_S$ & $\Sigma$ \\
   \hline 
   $U(1)_{X}$ & +2 & -2 & -3 & 1 & -1 & 1  & 5 & 0 & 0\\
   \hline 
   $Z_{2}$ & - & + & + & + & + & -  & + & + & -\\
   \hline  \hline
  \end{tabular}
 \caption{{\footnotesize Representations of Flipped $SU(5)$ with $U(1)_X$ and $Z_2$ charge assignments.}}
 \label{tab:U1-Z2}
\end{table}

 The complete superpotential which is renormalizable and invariant under $SU(5) \times U(1)_X$ and $Z_2$ matter parity, is given by
\bea\label{Eq:reheat-superpot}
W &=&  \kappa S \left( \mathbf{10}_{H}^{\alpha \beta} \,\,  \mathbf{\overline{10}}_{H{\alpha \beta}} - M^2 \right) - \frac{\mu}{2} \, S^2 + \frac{\lambda}{3} \, S^3 \,   + \, \lambda_1 \, \epsilon_{\alpha \beta \gamma \delta \zeta} \, \mathbf{10}_{H}^{\alpha \beta} \, \mathbf{10}_{H}^{\gamma \delta} \, \mathbf{5}_{H_d}^{\zeta} 
\nonumber \\
&& + \,
{ Y_u \,\mathbf{\bar{5}}_{H_u\alpha} \, \mathbf{10}_F^{\alpha\beta} \, \mathbf{\bar{5}}_{F\beta} } \, +\, Y_{d}\, \epsilon_{\alpha\beta\gamma\delta\zeta}\, \mathbf{10}_F^{\alpha\beta}\, \mathbf{10}_F^{\gamma\delta}\, \mathbf{5}_{H_d}^{\zeta}
\,+ \, Y_e \, \mathbf{5}_{H_d}^{\lambda} \, \mathbf{\bar{5}}_{F \lambda} \, \mathbf{1}_F\,
,
\eea
where the indices $\alpha,\beta,\cdots = 1,\cdots,5$ are the $SU(5)$ indices. 
We added $Z_2$ matter parity (Table \ref{tab:U1-Z2}), with charge assignments different from those in \cite{Kyae:2005nv,Ellis:2014xda,Gonzalo:2016gey}, in order to forbid undesirable terms as follows:
\begin{enumerate}
\item The bare mass term $\mu_H \, \mathbf{\bar 5}_{H_u} \mathbf{ 5}_{H_d}$ gives rise to the MSSM mixing term $\mu_H \, {H_u} \,{H_d}$, which is important to trigger the electroweak symmetry breaking in the MSSM and provide the mass for the Higgs superpartners. Indeed a new scale $\mu_H$ is not protected from  being a superheavy scale as the only mass scales in the theory are the GUT and the inflation high scales. Therefore $\mu_H$ should be put by hand to the SUSY scale to solve the hierarchy problem consistently. This is called the $\mu$-problem which is a kind of naturalness problem. Hence we forbid this term by $Z_2$ symmetry.

\item One may invoke a solution to the MSSM $\mu$-problem via the gauge invariant term $\lambda_{SH_5} S  \, \mathbf{\bar 5}_{H_u} \mathbf{ 5}_{H_d}$ which allows also for decay channels of the inflaton to the higgs. As indicated in \cite{Barbieri:1982eh,Dvali:1997uq}, the singlet $S$  can acquire a non-vanishing vev of order $\cal O$(TeV) if a large expectation value of $\mathbf{10}_{H}, \mathbf{\overline{10}}_{H}$ is triggered, hence SUSY breaking effects shift the vev of $S$ to a non-zero value of the same order as the SUSY breaking scale \cite{Barbieri:1982eh}. However, this solution is a lost cause. In fact the mixing between the inflaton $x$ and $\bar{\nu}^c_H$ gives rise to the coupling: %$\lambda_{SH_5}\,\sin\theta \, \bar{\nu}^c_H\, \left( \tilde{H}^0_1 \, \tilde{H}^0_2 \, + \,  \tilde{H}^+_2 \, \tilde{H}^-_1 \right)$
$\lambda_{SH_5}\,\sin\theta \, \bar{\nu}^c_H\, H_u \, H_d$ which results in a heavy mass term $\lambda_{SH_5}\,\sin\theta \,M$. If the reheating from inflaton decays constrains $\lambda_{SH_5}\sim 10^{-6}$, then $\sin\theta$ should be ${\cal O}(10^{-7})$ which is extremely small and not allowed by the values in the parameter space which are determined by inflation. Nonetheless, the former term is forbidden by $Z_2$ symmetry as well.

A natural solution to the MSSM $\mu$-problem is obtained
via Giudice-Masiero (GM) mechanism \cite{Giudice:1988yz}, or alternatively by putting $\mu_H=0$ at tree level and then generating it radiatively \cite{Hall:1983iz}.

\item The proposed $Z_2$ symmetry will forbid higher-dimensional baryon number violating operators such as $ \mathbf{10}_{F_i} \, \mathbf{10}_{F_j} \, \mathbf{10}_{F_k} \, \mathbf{\overline{5}}_{F_l}  $ and $\mathbf{10}_{F_i}  \, \mathbf{\overline{5}}_{F_j}  \, \mathbf{\overline{5}}_{F_k} \, \mathbf{1}_{F_l}$, where $i,j, \cdots$ are flavor indices. Therefore, proton decay will 
 occur via dimension six operators mediated by the supermassive GUT gauge bosons. Accordingly The proton life time predictions are consistent with the observation limits \cite{Ellis:2002vk,Dorsner:2004xx,Kyae:2005nv,Ellis:2017jcp}.

Moreover, terms such as $\mathbf{10}_{H} \, \mathbf{10}_{F_i} \, \mathbf{5}_{H_d}$ and $\mathbf{10}_{H} \, \mathbf{5}_{F_i} \, \mathbf{\bar{5}}_{H_u}$ are also forbidden by $Z_2$ symmetry. These terms may provide heavy masses to $d_i^c, L_i$ and $H_u$ via the terms 
$\langle \nu_H^c \rangle \,d_i^c \bar{D}^c$ and $\langle \nu_H^c \rangle \,L_i \, H_u$.\footnote{Here we list some advantages of   $Z_2$ symmetry which is proposed to allow decay channels for the inflaton to reheat the universe, generate Dirac mass terms for the fermions and neutrinos and prohibit mixing terms of $H_u, H_d$ that results in heavy masses for the Higgs. A complete phenomenology of flipped $SU(5)$ model is beyond the scope of this paper and one can find more details regarding the phenomenology in the literature.}
\end{enumerate}

Flipped $SU(5)$ provides a nice solution to the doublet-triplet splitting problem via the missing partner mechanism \cite{Antoniadis:1987dx, Kyae:2005nv}. As a matter of fact the the MSSM Higgs pair $H_u, H_d$
doesn't acquire masses throught the vev  $\langle \nu^c_H \rangle= \langle \bar{\nu}^c_H\rangle = M$ since the coupling term $\lambda_1 \mathbf{10}_{H}\,\mathbf{10}_{H} \, \mathbf{5}_{H_d}$  provides the components $\bar{D}^c$ of  $\mathbf{5}_{H_d}$, and the corresponding components $d_H^c$  of $\mathbf{10}_{H}$  with a superheavy mass scale proportional to the vev $M$. On the other hand, the coupling term $ \, \mathbf{\overline{10}}_{H} \, \mathbf{\overline{10}}_{H} \, \mathbf{\bar 5}_{H_u}$ is forbidden by $Z_2$ symmetry at tree level.  However it can be generated at the non-renormalizable level, hence providing the components ${D}^c$ of  $\mathbf{\bar 5}_{H_u}$, and the corresponding components $\bar{d}_H^c$  of $\mathbf{\overline{10}}_{H}$  with large masses. The components $Q_H $ and $\bar{Q}_H$ of $\mathbf{10}_{H}$ and $\mathbf{\overline{10}}_{H}$ will be absorbed by the gauge bosons to render them massive.
%$D^c$ of  $\mathbf{\bar 5}_{H_u}$ and and $\bar{d}_H^c$ of  $\mathbf{\overline{10}}_{H}$,

Now we discuss the terms which are relevant to the reheating.
 The breaking of R-symmetry allows the non-renormalizable term $\dfrac{\lambda_2}{M_{P}} \,\mathbf{10}_{F}  \, \mathbf{10}_{F} \, \mathbf{\overline{10}}_{H} \, \mathbf{\overline{10}}_{H}$ which is invariant under FGUT gauge group. This term gives rise to right-handed neutrino masses and allows for decay channels for the inflaton to right-handed neutrinos ${\nu}^c$ and sneutrinos $\tilde{\nu}^{c}$, via the mixing with the FGUT higgs. Hence for $M \sim 10^{15} -10^{16}$ GeV and $\lambda_2 \sim {\cal O}(10^{-4}-1)$, the right-handed neutrino mass is $M_{\nu^c}\sim {\cal O}(10^{8}-10^{14})$ GeV which is suitable for generating the tiny neutrino masses via type I seesaw mechanism.
In that respect, the interaction Lagrangian responsible for the inflaton decay is given by
\bea\label{Eq:reheat1}
{\cal L}_{\text int}=  \,
\dfrac{\lambda_2\, M}{M_{P}}\, \bar{\nu}_H^c \, \nu^c\, \nu^c \, + \, 
\left( \dfrac{2\kappa \lambda_2 M^2}{M_{P}} \, S \,\tilde{\nu}^{c\,*} \,\tilde{\nu}^{c\,*}\, + \, h.c. \right).
\eea
After diagonalizing the mass matrix (\ref{Eq:FHImass}), we rewrite the Lagrangian (\ref{Eq:reheat1}) in terms of the physical states $x',\alpha'$
\bea\label{Eq:reheat2}
{\cal L}_{\text int} &=& x' \left[\left( \, \dfrac{2\kappa \lambda_2 \cos \theta\sqrt{\tau_0} M^2}{M_{P}} \, \,\tilde{\nu}^{c\,*} \,\tilde{\nu}^{c\,*}\, +\, h.c\right) \, + \,  \dfrac{\lambda_2\, \sin\theta\, M}{M_{P}} \, \nu^c\, \nu^c \right]\, \nonumber \\
&& + \alpha' \left[\dfrac{\lambda_4\, M}{M_{P}}\,  \cos \theta \, \nu^c\, \nu^c
-\, \left( \dfrac{2\kappa \lambda_2 \sin \theta\sqrt{\tau_0} M^2}{M_{P}} \,\tilde{\nu}^{c\,*}\,\tilde{\nu}^{c\,*}\,\,  + \, h.c.\right)\right]
\eea 
The reheating temperature is given by \cite{Lazarides:1996dv,Lazarides:2001zd}
\bea\label{Eq:TR}
T_R \approx \frac{ (8\pi)^{1/4}}{7}\left( \Gamma_{x'}\,  M_p\right)^{1/2},
\eea
where $\Gamma_{x'}$ is the total decay width of the inflaton field which is given by 
\be
\Gamma_{x'}=\Gamma_{x'\to \,\tilde{\nu}^{c\,*}\,\tilde{\nu}^{c\,*}}+\Gamma_{x'\to \, \nu^c\, \nu^c}= 
\frac{m_{x'}}{8\pi} \left[
\left( \, \dfrac{2\kappa \lambda_2 \cos \theta\sqrt{\tau_0} M^2}{m_{x'}\, M_{P}} \right)^2\,+\, \left(\dfrac{\lambda_2\, \sin\theta\, M}{M_{P}} \right)^2 \, 
\right]
\,.
\ee
An upper bound on the reheating temperature arises from cosmological constraints such as the gravitino overproduction problem \cite{Ellis:1984eq,Ellis:1984er,Moroi:1993mb,Kawasaki:2004yh, Antusch:2010mv},  and it is given by \cite{Antusch:2010mv}
\begin{equation}
 T_R < 10^{7} - 10^{10} \text{ GeV}.
\end{equation}
\begin{figure}[t!]
	\centering
	\includegraphics[width=0.6\textwidth]{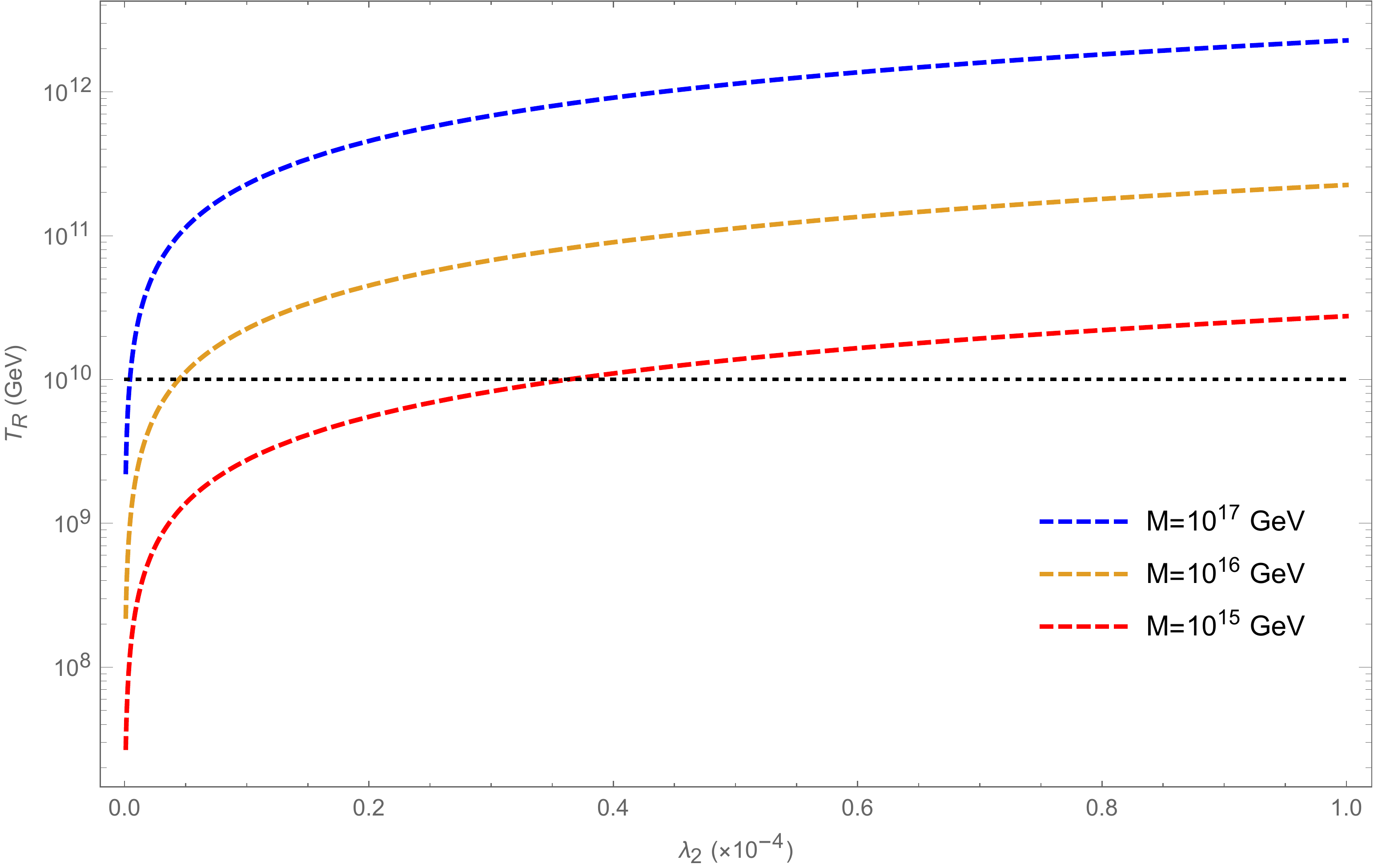}
	\caption{A logarithmic plot for the variation of the reheating temperature versus the change in the coupling $\lambda_2$ for different values of the GUT scale $M=10^{17},10^{16},10^{15}$ GeV.
	\label{fig:Reheat2}}
\end{figure}	
Fig. \ref{fig:Reheat2} depicts the relation between the reheating temperature and the coupling $\lambda_2$ according to Eq. (\ref{Eq:TR}). The GUT scale is fixed with three different values $M=10^{17},10^{16},10^{15}$ GeV. We have fixed the other parameters with values consistent with the inflation observables: $\hat{\mu} \sim \lambda\sim 10^{-5}$, $\kappa \sim 10^{-2}$. The black dotted line gives the cosmological constrain on the reheating temperature. The figure shows that a reheating temperature consistent with the observation bound prefers the GUT scale $M=10^{15}$ GeV and $\lambda_2\lesssim 0.35 \times 10^{-4}$.
This produces right-handed neutrino masses $M_{\nu^c}\sim 10^{8}$ GeV and inflaton mass $m_{x'} \sim 10^{13}$ GeV, hence the decay is kinematically allowed. The tiny neutrino masses are generated via the seesaw mechanism: $m_\nu = \dfrac{Y_u^2 v^2\sin^2\beta}{M_{\nu^c}}$, where $v^2= \langle H_u^0\rangle^2+\langle H_d^0\rangle^2$ and $\beta$ is the mixing angle in the MSSM neutral Higgs sector. As hinted by global fit scans \cite{deVries:2015hva}, $10 \lesssim tan\beta  \lesssim 50$, and hence $\sin\beta\sim {\cal O}(1)$. Therefore, in order to have neutrino masses of the order of eV, $Y_u \lesssim {\cal O}(10^{-3})$, which is typically the same order as the first and second generations. In deed, we may consider the different flavors interactions $\dfrac{\lambda_2^{ij}}{M_{P}} \,\mathbf{10}_{Fi}  \, \mathbf{10}_{Fj} \, \mathbf{\overline{10}}_{H} \, \mathbf{\overline{10}}_{H}$, hence for up sector Yukawa couplings $Y_u^{ij}\sim {\text diag}(10^{-5},7\times 10^{-3},1)$,  $\lambda_2^{ij} \lesssim {\text diag}(10^{-9}, 10^{-4},1)$. 
Therefore the inflaton should decay only to the first or second generations in the neutrino sector, as contributions from the the third generation would drive the reheating temperature to values exceeding the the above constrain. We may assume the existence of a flavour violating
sector at high energies, without specific details, that prevents the decay to the third generation.

%%%%%%%%%%%%%%%%%%%%%%%%%%%%%%%%%%%%%%%%%%%%%%%%%%%%%%%%%%%%%%%%%%%%%%%%%%%%%%%%%%%%%%%%%%%%%%%%%%%
\section{Conclusions}
\label{sec:conclusions}
%%%%%%%%%%%%%%%%%%%%%%%%%%%%%%
In this paper, we have proposed a scenario for hybrid inflation with a maximal breaking of R-symmetry, in no-scale supergravity context. In that respect, we have studied the FHI model and found a region in the parameter space in which the effective potential is asymptotically flat, which is not studied in \cite{Romao:2017uwa}. We have treated the dynamics of all fields in full detail and realized the waterfall phase. Moreover, we have discussed the case of adding FI D-term which results in an effective potential similar to the ENO model but shifted.

The question of moduli stabilization and their backreaction as well as SUSY breaking has been discussed. It has been emphasised that inflation trajectory will not be affected in specific type of strong moduli stabilization proposed by Ellis et al. 

Finally, the reheating phase has been studied in the context of flipped GUT scenario. We have stressed on the role of the associated $Z_2$ symmetry in low energy phenomenology and in allowing for decay channels for the inflaton in connection to the neutrino masses. We emphasised that the calculated reheating temperature prefers GUT symmetry breaking scale $M\sim 10^{15}$ GeV, in order to be consistent with the cosmological constraints. Moreover the latter constrain implies that the inflaton should decay only to the first or second generations in the neutrino sector, while the third generation should decouple.

%%%%%%%%%%%%%%%%%%%%%%%%%%%%%%%%%%%%%%%%%%%%%%%%%%%%%%%%%%%%%%%%%%%%%%%%%%%%%%%%%%%%%%%%%%%%%%%%%%%
\section*{Acknowledgments}
The work of A. M. is partially supported by the Science, Technology and Innovation Funding Authority (STDF) under grant No. (33495) YRG.
%%%%%%%%%%%%%%%%%%%%%%%%%%%%%%%%%%%%%%%%%%%%%%%%%%%%%%%%%%%%%%%%%%%%%%%%%%%%%%%%
\appendix

\section{No-scale FHI observables}
\label{App:starolike}
The crossing horizon value of the inflaton $x_*$ as a function in $N,b$ is given by

\bea
x_* \simeq \,\sqrt{\frac{3}{2}} \, \log \left[\frac{(1-b) \left(4 N+5 -3 \log \left(\frac{5-5 b}{3 b+3}\right)\right)}{3 (b+1)}\right]
\eea
One can find relations between $n_s,N$ and $r,N$ as follows
\bea
N &=& \frac{1}{4} \left[8 \sqrt{\frac{ 3}{r}}+3\log (5)-3 \log \left(8 \sqrt{\frac{ 3}{r}}+3\right)-2\right]\nonumber\\
N &=& -\frac{4 \sqrt{4-3 n_s}-n_s (\log (125)-2)+3 (n_s-1) \log \left(\frac{3 n_s-4 \sqrt{4-3 n_s}-7}{n_s-1}\right)+2+\log (125)}{4 (n_s-1)}
\eea

%%%%%%%%%%%%%%%%%%%%%%%%%%%%%%%%%%%%%%%%%%%%%%%%%%%%%%%%%%%%%%%%%%%%%%%%%%%%%%%%%
\section{Mass matrices}
\label{sec:Mass}
For FHI and in the basis $(x,\alpha,y,\beta)$, the target space metric $g_{ij}$ is diagonal at the minimum, and is given by 
\bea
g_{ij}\Bigg|_{\text min} = \dfrac{3 \tau_0}{3 \tau_0- M^2} diag\left[ 1 , \frac{3}{3 \tau_0- M^2} , 1  , \frac{3}{3 \tau_0- M^2} \right]
\eea
Therefore the mass squared matrix in the basis $(x,\alpha,y,\beta)$, ($\alpha={\nu}_H^c$), is given by

\bea\label{Eq:FHImass}
m= \left(
\begin{array}{cc}
 {\cal M} & 0 \\
   0 & {\cal M} \\
\end{array}
\right)\,\,\,,
%\eea
%
\text{with}\,\,\,
%
%\bea
{\cal M}=
\left(
\begin{array}{cc}
 \frac{27 \tau_0^2 \left(2 M^2 \kappa ^2+\mu ^2\right)}{2\left(3 \tau_0- M^2\right)^3} & -\frac{27 \sqrt{6} M \tau_0^{3/2} \kappa  \mu }{2\left(3 \tau_0- M^2\right)^{7/2}}  \\
 -\frac{27 \sqrt{6} M \tau_0^{3/2} \kappa  \mu }{2\left(3 \tau_0- M^2\right)^{7/2}}& \frac{81 \tau_0\, M^2 \kappa ^2}{\left(3 \tau_0- M^2\right)^4}\\
\end{array}
\right)
\eea

For FDHI, in the basis $(x,\alpha_1,y,\beta_1,\alpha_2,\beta_2)$, the target space metric $g_{ij}$ is diagonal at the minimum, and is given by 
\bea
g_{ij}\Bigg|_{\text min}= diag\left[ 1+\dfrac{\xi}{3} , \frac{3+\xi}{6\tau_0} , 1+\dfrac{\xi}{3} , \frac{3+\xi}{6 \tau_0}, \dfrac{(3+\xi)^2}{18 \tau_0} , \dfrac{(3+\xi)^2}{18 \tau_0} \right]. \nonumber
\eea
Hence the masses are given by 
\bea
m_{\alpha_2}^2= \frac{g^2 \xi  (\xi +3)^5}{486 \tau_0^2}\,, \hspace{1cm} m_{\beta_2}^2= 0\,,
\eea
and the mass squared matrix of the other fields $(x,\alpha_1,y,\beta_1)$ is given by 
\bea
m^2= \left(
\begin{array}{cc}
 {\cal M}^2 & 0 \\
0 & {\cal M}^2 \\
\end{array}
\right)\,\,\,,
%\eea
%
\text{with}\,\,\,
%
%\bea
{\cal M}=
\left(
\begin{array}{cc}
 \frac{(\xi +3)^2 \left(6 \tau_0 \xi  \kappa ^2+\mu ^2 (\xi +3)\right)}{54 \tau_0} & \frac{ 2 \tau_0 \kappa  \mu  \left(\frac{\xi  (\xi +3)}{2 \tau_0}\right)^{5/2}}{9 \sqrt{3} \xi ^2} \\
 \frac{ 2 \tau_0 \kappa  \mu  \left(\frac{\xi  (\xi +3)}{2 \tau_0}\right)^{5/2}}{9 \sqrt{3} \xi ^2} & \frac{\kappa ^2 \xi  (\xi +3)^2}{36 \tau_0^2} \\
\end{array}
\right)
\eea
%%%%%%%%%%%%%%%%%%%%%%%%%%%%%%%%%%%%%%%%%%%%%%%%%%%%%%%%%%%%%%%%%%%%%%%%%%%%%%%%
%%%%%%%%%%%%%%%%%%%%%%%%%%%%%%%%%%%%%%%%%%%%%%%%%%%%%%%%%%%%%%%%%%%%%%%%%%%%%%%%%%%%%%


\begin{thebibliography}{99}
%

%\cite{Akrami:2018odb}
\bibitem{Akrami:2018odb}
Y.~Akrami \textit{et al.} [Planck],
``Planck 2018 results. X. Constraints on inflation,''
[arXiv:1807.06211 [astro-ph.CO]].
%753 citations counted in INSPIRE as of 12 May 2020

%\cite{Kawasaki:2000yn}
\bibitem{Kawasaki:2000yn}
M.~Kawasaki, M.~Yamaguchi and T.~Yanagida,
``Natural chaotic inflation in supergravity,''
Phys. Rev. Lett. \textbf{85} (2000), 3572-3575
doi:10.1103/PhysRevLett.85.3572
[arXiv:hep-ph/0004243 [hep-ph]].
%435 citations counted in INSPIRE as of 14 Sep 2020

%\cite{Yamaguchi:2000vm}
\bibitem{Yamaguchi:2000vm}
M.~Yamaguchi and J.~Yokoyama,
``New inflation in supergravity with a chaotic initial condition,''
Phys. Rev. D \textbf{63} (2001), 043506
doi:10.1103/PhysRevD.63.043506
[arXiv:hep-ph/0007021 [hep-ph]].
%61 citations counted in INSPIRE as of 14 Sep 2020

%\cite{Brax:2005jv}
\bibitem{Brax:2005jv}
P.~Brax and J.~Martin,
``Shift symmetry and inflation in supergravity,''
Phys. Rev. D \textbf{72} (2005), 023518
doi:10.1103/PhysRevD.72.023518
[arXiv:hep-th/0504168 [hep-th]].
%52 citations counted in INSPIRE as of 14 Sep 2020
%
%\cite{Kallosh:2010ug}
\bibitem{Kallosh:2010ug}
R.~Kallosh and A.~Linde,
``New models of chaotic inflation in supergravity,''
JCAP \textbf{11}, 011 (2010)
%doi:10.1088/1475-7516/2010/11/011
[arXiv:1008.3375 [hep-th]].
%224 citations counted in INSPIRE as of 05 Dec 2020
%
%\cite{Heurtier:2015ima}
\bibitem{Heurtier:2015ima}
L.~Heurtier, S.~Khalil and A.~Moursy,
%``Single Field Inflation in Supergravity with a $U(1)$ Gauge Symmetry,''
JCAP \textbf{10} (2015), 045
doi:10.1088/1475-7516/2015/10/045
[arXiv:1505.07366 [hep-ph]].
%8 citations counted in INSPIRE as of 14 Sep 2020

%\cite{Gonzalo:2016gey}
\bibitem{Gonzalo:2016gey}
T.~E.~Gonzalo, L.~Heurtier and A.~Moursy,
``Sneutrino driven GUT Inflation in Supergravity,''
JHEP \textbf{06} (2017), 109
%doi:10.1007/JHEP06(2017)109
[arXiv:1609.09396 [hep-th]].
%6 citations counted in INSPIRE as of 15 Aug 2020

%\cite{Starobinsky:1980te}
\bibitem{Starobinsky:1980te} 
  A.~A.~Starobinsky,
  ``A New Type of Isotropic Cosmological Models Without Singularity,''
  Phys.\ Lett.\  {\bf 91B}, 99 (1980);
  %\cite{Mukhanov:1981xt}
%\bibitem{Mukhanov:1981xt} 
  V.~F.~Mukhanov and G.~V.~Chibisov,
 ``Quantum Fluctuations and a Nonsingular Universe,''
  JETP Lett.\  {\bf 33}, 532 (1981)
  [Pisma Zh.\ Eksp.\ Teor.\ Fiz.\  {\bf 33}, 549 (1981)];
  %\cite{Starobinsky:1983zz}
%\bibitem{Starobinsky:1983zz} 
  A.~A.~Starobinsky,
  ``The Perturbation Spectrum Evolving from a Nonsingular Initially De-Sitter Cosmology and the Microwave Background Anisotropy,''
  Sov.\ Astron.\ Lett.\  {\bf 9}, 302 (1983).
%


%\cite{Linde:1993cn}
\bibitem{Linde:1993cn}
A.~D.~Linde,
``Hybrid inflation,''
Phys. Rev. D \textbf{49} (1994), 748-754
%doi:10.1103/PhysRevD.49.748
[arXiv:astro-ph/9307002 [astro-ph]].
%1241 citations counted in INSPIRE as of 29 Jul 2020
%

%\cite{Dvali:1994ms}
\bibitem{Dvali:1994ms}
  G.~R.~Dvali, Q.~Shafi and R.~K.~Schaefer,
  ``Large scale structure and supersymmetric inflation without fine tuning,''
  Phys.\ Rev.\ Lett.\  {\bf 73}, 1886 (1994).
  [hep-ph/9406319].
  
%\cite{BasteroGil:2006cm}
\bibitem{BasteroGil:2006cm}
M.~Bastero-Gil, S.~F.~King and Q.~Shafi,
``Supersymmetric Hybrid Inflation with Non-Minimal Kahler potential,''
Phys. Lett. B \textbf{651} (2007), 345-351
%doi:10.1016/j.physletb.2006.06.085
[arXiv:hep-ph/0604198 [hep-ph]].
%118 citations counted in INSPIRE as of 14 Sep 2020

%\cite{Dvali:1997uq}
\bibitem{Dvali:1997uq}
G.~R.~Dvali, G.~Lazarides and Q.~Shafi,
``Mu problem and hybrid inflation in supersymmetric SU(2)-L x SU(2)-R x U(1)-(B-L),''
Phys. Lett. B \textbf{424} (1998), 259-264
%doi:10.1016/S0370-2693(98)00145-2
[arXiv:hep-ph/9710314 [hep-ph]].
%120 citations counted in INSPIRE as of 21 Aug 2020

%\cite{Lazarides:1998iq}
\bibitem{Lazarides:1998iq}
G.~Lazarides and Q.~Shafi,
``R symmetry in minimal supersymmetry standard model and beyond with several consequences,''
Phys. Rev. D \textbf{58} (1998), 071702
%doi:10.1103/PhysRevD.58.071702
[arXiv:hep-ph/9803397 [hep-ph]].
%130 citations counted in INSPIRE as of 18 Sep 2020

%\cite{Hall:1983iz}
\bibitem{Hall:1983iz}
L.~J.~Hall, J.~D.~Lykken and S.~Weinberg,
``Supergravity as the Messenger of Supersymmetry Breaking,''
Phys. Rev. D \textbf{27} (1983), 2359-2378
%doi:10.1103/PhysRevD.27.2359
%1565 citations counted in INSPIRE as of 27 Sep 2020

%\cite{Giudice:1988yz}
\bibitem{Giudice:1988yz}
G.~F.~Giudice and A.~Masiero,
``A Natural Solution to the mu Problem in Supergravity Theories,''
Phys. Lett. B \textbf{206} (1988), 480-484
%doi:10.1016/0370-2693(88)91613-9
%978 citations counted in INSPIRE as of 02 Sep 2020

%\cite{Nelson:1993nf}
\bibitem{Nelson:1993nf}
A.~E.~Nelson and N.~Seiberg,
``R symmetry breaking versus supersymmetry breaking,''
Nucl. Phys. B \textbf{416} (1994), 46-62
%doi:10.1016/0550-3213(94)90577-0
[arXiv:hep-ph/9309299 [hep-ph]].
%350 citations counted in INSPIRE as of 28 Sep 2020

%\cite{Civiletti:2013cra}
\bibitem{Civiletti:2013cra}
M.~Civiletti, M.~Ur Rehman, E.~Sabo, Q.~Shafi and J.~Wickman,
``R-symmetry breaking in supersymmetric hybrid inflation,''
Phys. Rev. D \textbf{88} (2013) no.10, 103514
%doi:10.1103/PhysRevD.88.103514
[arXiv:1303.3602 [hep-ph]].
%7 citations counted in INSPIRE as of 15 Aug 2020


%\cite{Khalil:2018iip}
\bibitem{Khalil:2018iip}
S.~Khalil, A.~Moursy, A.~K.~Saha and A.~Sil,
``U(1)R inspired inflation model in no-scale supergravity,''
Phys. Rev. D \textbf{99} (2019) no.9, 095022
%doi:10.1103/PhysRevD.99.095022
[arXiv:1810.06408 [hep-ph]].
%5 citations counted in INSPIRE as of 15 Aug 2020

%\cite{Schmitz:2016kyr}
\bibitem{Schmitz:2016kyr}
K.~Schmitz and T.~T.~Yanagida,
``Dynamical supersymmetry breaking and late-time R symmetry breaking as the origin of cosmic inflation,''
Phys. Rev. D \textbf{94}, no.7, 074021 (2016)
%doi:10.1103/PhysRevD.94.074021
[arXiv:1604.04911 [hep-ph]].
%21 citations counted in INSPIRE as of 07 Dec 2020
%\cite{Domcke:2017xvu}
%\bibitem{Domcke:2017xvu}
V.~Domcke and K.~Schmitz,
``Unified model of D-term inflation,''
Phys. Rev. D \textbf{95}, no.7, 075020 (2017)
%doi:10.1103/PhysRevD.95.075020
[arXiv:1702.02173 [hep-ph]].
%21 citations counted in INSPIRE as of 07 Dec 2020
%\cite{Domcke:2017rzu}
%\bibitem{Domcke:2017rzu}
V.~Domcke and K.~Schmitz,
``Inflation from High-Scale Supersymmetry Breaking,''
Phys. Rev. D \textbf{97}, no.11, 115025 (2018)
%doi:10.1103/PhysRevD.97.115025
[arXiv:1712.08121 [hep-ph]].
%16 citations counted in INSPIRE as of 07 Dec 2020 


%\cite{Kachru:2003aw}
\bibitem{Kachru:2003aw} 
  S.~Kachru, R.~Kallosh, A.~D.~Linde and S.~P.~Trivedi,
  ``De Sitter vacua in string theory,''
  Phys.\ Rev.\ D {\bf 68}, 046005 (2003),
%  doi:10.1103/PhysRevD.68.046005
  [hep-th/0301240].
%
%\cite{Balasubramanian:2005zx}
\bibitem{Balasubramanian:2005zx} 
  V.~Balasubramanian, P.~Berglund, J.~P.~Conlon and F.~Quevedo,
  ``Systematics of moduli stabilisation in Calabi-Yau flux compactifications,''
  JHEP {\bf 0503}, 007 (2005)
%  doi:10.1088/1126-6708/2005/03/007
  [hep-th/0502058]; 
  %\cite{Balasubramanian:2005zx}
%\bibitem{Balasubramanian:2005zx} 
  V.~Balasubramanian, P.~Berglund, J.~P.~Conlon and F.~Quevedo,
  ``Systematics of moduli stabilisation in Calabi-Yau flux compactifications,''
  JHEP {\bf 0503}, 007 (2005)
%  doi:10.1088/1126-6708/2005/03/007
  [hep-th/0502058].
%

%\cite{Cremmer:1983bf}
\bibitem{Cremmer:1983bf}
E.~Cremmer, S.~Ferrara, C.~Kounnas and D.~V.~Nanopoulos,
``Naturally Vanishing Cosmological Constant in N=1 Supergravity,''
Phys. Lett. B \textbf{133} (1983), 61.
%doi:10.1016/0370-2693(83)90106-5
%798 citations counted in INSPIRE as of 19 Sep 2020

%\cite{Ellis:1984bm}
\bibitem{Ellis:1984bm}
J.~R.~Ellis, C.~Kounnas and D.~V.~Nanopoulos,
``No Scale Supersymmetric Guts,''
Nucl. Phys. B \textbf{247} (1984), 373-395.
%doi:10.1016/0550-3213(84)90555-8
%597 citations counted in INSPIRE as of 19 Sep 2020

%\cite{Cecotti:1987sa}
\bibitem{Cecotti:1987sa}
S.~Cecotti,
``HIGHER DERIVATIVE SUPERGRAVITY IS EQUIVALENT TO STANDARD SUPERGRAVITY COUPLED TO MATTER. 1.,''
Phys. Lett. B \textbf{190} (1987), 86-92
%doi:10.1016/0370-2693(87)90844-6
%162 citations counted in INSPIRE as of 19 Sep 2020

%\cite{Kallosh:2013lkr}
\bibitem{Kallosh:2013lkr}
R.~Kallosh and A.~Linde,
``Superconformal generalizations of the Starobinsky model,''
JCAP \textbf{06} (2013), 028
%doi:10.1088/1475-7516/2013/06/028
[arXiv:1306.3214 [hep-th]].
%236 citations counted in INSPIRE as of 19 Sep 2020

%\cite{Ellis:2013xoa}
\bibitem{Ellis:2013xoa} 
  J.~Ellis, D.~V.~Nanopoulos and K.~A.~Olive,
  ``No-Scale Supergravity Realization of the Starobinsky Model of Inflation,''
  Phys.\ Rev.\ Lett.\  {\bf 111}, 111301 (2013)
  Erratum: [Phys.\ Rev.\ Lett.\  {\bf 111}, no. 12, 129902 (2013)]
%  doi:10.1103/PhysRevLett.111.129902, 10.1103/PhysRevLett.111.111301
  [arXiv:1305.1247 [hep-th]].
%


\bibitem{Romao:2017uwa} 
  M.~C.~Romao and S.~F.~King,
  ``Starobinsky-like inflation in no-scale supergravity Wess-Zumino model with Polonyi term,''
  JHEP {\bf 1707}, 033 (2017)
%  doi:10.1007/JHEP07(2017)033
  [arXiv:1703.08333 [hep-ph]];
%\cite{King:2019omb}
%\bibitem{King:2019omb}
S.~F.~King and E.~Perdomo,
``Starobinsky-like inflation and soft-SUSY breaking,''
JHEP \textbf{05}, 211 (2019)
%doi:10.1007/JHEP05(2019)211
[arXiv:1903.08448 [hep-ph]].
%5 citations counted in INSPIRE as of 27 Jul 2020 

%\cite{Coleman:1973jx}
\bibitem{Coleman:1973jx}
S.~R.~Coleman and E.~J.~Weinberg,
``Radiative Corrections as the Origin of Spontaneous Symmetry Breaking,''
Phys. Rev. D \textbf{7} (1973), 1888-1910
%doi:10.1103/PhysRevD.7.1888
%4477 citations counted in INSPIRE as of 28 Sep 2020

%\cite{Binetruy:1996xj}
\bibitem{Binetruy:1996xj}
P.~Binetruy and G.~R.~Dvali,
``D term inflation,''
Phys. Lett. B \textbf{388}, 241-246 (1996)
%doi:10.1016/S0370-2693(96)01083-0
[arXiv:hep-ph/9606342 [hep-ph]].
%446 citations counted in INSPIRE as of 27 Jul 2020

%\cite{Halyo:1996pp}
\bibitem{Halyo:1996pp}
E.~Halyo,
``Hybrid inflation from supergravity D terms,''
Phys. Lett. B \textbf{387} (1996), 43-47
%doi:10.1016/0370-2693(96)01001-5
[arXiv:hep-ph/9606423 [hep-ph]].
%352 citations counted in INSPIRE as of 30 Jul 2020

%\cite{Binetruy:2004hh}
\bibitem{Binetruy:2004hh}
P.~Binetruy, G.~Dvali, R.~Kallosh and A.~Van Proeyen,
``Fayet-Iliopoulos terms in supergravity and cosmology,''
Class. Quant. Grav. \textbf{21} (2004), 3137-3170
%doi:10.1088/0264-9381/21/13/005
[arXiv:hep-th/0402046 [hep-th]].
%197 citations counted in INSPIRE as of 30 Jul 2020

%\cite{Ellis:2013nxa}
\bibitem{Ellis:2013nxa}
J.~Ellis, D.~V.~Nanopoulos and K.~A.~Olive,
``Starobinsky-like Inflationary Models as Avatars of No-Scale Supergravity,''
JCAP \textbf{10} (2013), 009 
%doi:10.1088/1475-7516/2013/10/009
[arXiv:1307.3537 [hep-th]].
%181 citations counted in INSPIRE as of 30 Jul 2020

%\cite{Ellis:1984bs}
\bibitem{Ellis:1984bs}
J.~R.~Ellis, C.~Kounnas and D.~V.~Nanopoulos,
``No Scale Supergravity Models with a Planck Mass Gravitino,''
Phys. Lett. B \textbf{143} (1984), 410-414
%doi:10.1016/0370-2693(84)91492-8
%167 citations counted in INSPIRE as of 30 Jul 2020

%\cite{Ellis:2020xmk}
\bibitem{Ellis:2020xmk}
J.~Ellis, D.~V.~Nanopoulos, K.~A.~Olive and S.~Verner,
``Phenomenology and Cosmology of No-Scale Attractor Models of Inflation,''
[arXiv:2004.00643 [hep-ph]].
%2 citations counted in INSPIRE as of 12 Aug 2020

%%%%%%%%%%%%%%%%%%%%%%%%%%%%%%%%%%% Moduli backreaction %%%%%%%%%%%%%%%%%%%%%%%%%%%%%%%%%%%%%%%
%\cite{Brax:2006ay}
\bibitem{Brax:2006ay}
P.~Brax, C.~van de Bruck, A.~C.~Davis and S.~C.~Davis,
``Coupling hybrid inflation to moduli,''
JCAP \textbf{09} (2006), 012
%doi:10.1088/1475-7516/2006/09/012
[arXiv:hep-th/0606140 [hep-th]];
%21 citations counted in INSPIRE as of 30 Jul 2020
%
%\cite{Brax:2006yq}
\bibitem{Brax:2006yq}
P.~Brax, C.~van de Bruck, A.~C.~Davis, S.~C.~Davis, R.~Jeannerot and M.~Postma,
``Moduli corrections to D-term inflation,''
JCAP \textbf{01} (2007), 026
%doi:10.1088/1475-7516/2007/01/026
[arXiv:hep-th/0610195 [hep-th]].
%24 citations counted in INSPIRE as of 13 Aug 2020

%\cite{Davis:2008sa}
%\bibitem{Davis:2008sa}
S.~C.~Davis and M.~Postma,
``Successfully combining SUGRA hybrid inflation and moduli stabilisation,''
JCAP \textbf{04} (2008), 022
%doi:10.1088/1475-7516/2008/04/022
[arXiv:0801.2116 [hep-th]];
%34 citations counted in INSPIRE as of 30 Jul 2020
%
%\cite{Mooij:2010cs}
%\bibitem{Mooij:2010cs}
S.~Mooij and M.~Postma,
``Hybrid inflation with moduli stabilization and low scale supersymmetry breaking,''
JCAP \textbf{06} (2010), 012
%doi:10.1088/1475-7516/2010/06/012
[arXiv:1001.0664 [hep-ph]];
%10 citations counted in INSPIRE as of 30 Jul 202
%


%\cite{Linde:2011ja}
\bibitem{Linde:2011ja}
A.~Linde, Y.~Mambrini and K.~A.~Olive,
``Supersymmetry Breaking due to Moduli Stabilization in String Theory,''
Phys. Rev. D \textbf{85} (2012), 066005
%doi:10.1103/PhysRevD.85.066005
[arXiv:1111.1465 [hep-th]].
%50 citations counted in INSPIRE as of 30 Jul 2020

%\cite{Buchmuller:2013uta}
\bibitem{Buchmuller:2013uta}
W.~Buchmüller, V.~Domcke and C.~Wieck,
``No-scale D-term inflation with stabilized moduli,''
Phys. Lett. B \textbf{730} (2014), 155-160
%doi:10.1016/j.physletb.2014.01.040
[arXiv:1309.3122 [hep-th]].
%24 citations counted in INSPIRE as of 30 Jul 2020


%\cite{Buchmuller:2014vda}
\bibitem{Buchmuller:2014vda}
W.~Buchmuller, C.~Wieck and M.~W.~Winkler,
``Supersymmetric Moduli Stabilization and High-Scale Inflation,''
Phys. Lett. B \textbf{736} (2014), 237-240
%doi:10.1016/j.physletb.2014.07.024
[arXiv:1404.2275 [hep-th]].
%43 citations counted in INSPIRE as of 30 Jul 2020

%\cite{Buchmuller:2014pla}
\bibitem{Buchmuller:2014pla}
W.~Buchmuller, E.~Dudas, L.~Heurtier and C.~Wieck,
``Large-Field Inflation and Supersymmetry Breaking,''
JHEP \textbf{09} (2014), 053
%doi:10.1007/JHEP09(2014)053
[arXiv:1407.0253 [hep-th]].
%54 citations counted in INSPIRE as of 30 Jul 2020

%\cite{Buchmuller:2015oma}
\bibitem{Buchmuller:2015oma}
W.~Buchmuller, E.~Dudas, L.~Heurtier, A.~Westphal, C.~Wieck and M.~W.~Winkler,
``Challenges for Large-Field Inflation and Moduli Stabilization,''
JHEP \textbf{04} (2015), 058
doi:10.1007/JHEP04(2015)058
[arXiv:1501.05812 [hep-th]].
%72 citations counted in INSPIRE as of 30 Jul 2020
%

%\cite{Davis:2008fv}
\bibitem{Davis:2008fv}
S.~C.~Davis and M.~Postma,
``SUGRA chaotic inflation and moduli stabilisation,''
JCAP \textbf{03} (2008), 015
%doi:10.1088/1475-7516/2008/03/015
[arXiv:0801.4696 [hep-ph]].
%57 citations counted in INSPIRE as of 30 Jul 2020

%\cite{Wieck:2014xxa}
\bibitem{Wieck:2014xxa}
C.~Wieck and M.~W.~Winkler,
``Inflation with Fayet-Iliopoulos Terms,''
Phys. Rev. D \textbf{90} (2014) no.10, 103507
%doi:10.1103/PhysRevD.90.103507
[arXiv:1408.2826 [hep-th]].
%27 citations counted in INSPIRE as of 30 Jul 2020

%\cite{Dudas:2015lga}
\bibitem{Dudas:2015lga}
E.~Dudas and C.~Wieck,
``Moduli backreaction and supersymmetry breaking in string-inspired inflation models,''
JHEP \textbf{10} (2015), 062
%doi:10.1007/JHEP10(2015)062
[arXiv:1506.01253 [hep-th]].
%37 citations counted in INSPIRE as of 30 Jul 2020 

%%%%%%%%%%%%%%%%%%%%%%%%%%%%%%%%  Reheating %%%%%%%%%%%%%%%%%%%%%%%%%%%%%%%%%%%%%%%%%%%%%%%%%%%%%%%%%%%%%%
%\cite{tHooft:1974kcl}
\bibitem{tHooft:1974kcl}
G.~'t Hooft,
``Magnetic Monopoles in Unified Gauge Theories,''
Nucl. Phys. B \textbf{79} (1974), 276-284.
%doi:10.1016/0550-3213(74)90486-6
%3022 citations counted in INSPIRE as of 18 Aug 2020

%\cite{Ellis:2014xda}
\bibitem{Ellis:2014xda}
J.~Ellis, T.~E.~Gonzalo, J.~Harz and W.~C.~Huang,
``Flipped GUT Inflation,''
JCAP \textbf{03} (2015), 039
%doi:10.1088/1475-7516/2015/03/039
[arXiv:1412.1460 [hep-ph]].
%16 citations counted in INSPIRE as of 22 Aug 2020

\bibitem{Antoniadis:1987dx}
I.~Antoniadis, J.~R.~Ellis, J.~S.~Hagelin and D.~V.~Nanopoulos,
``Supersymmetric Flipped SU(5) Revitalized,''
Phys. Lett. B \textbf{194} (1987), 231-235;
%doi:10.1016/0370-2693(87)90533-8
%562 citations counted in INSPIRE as of 21 Aug 2020;
%\cite{Masiero:1982fe}
%\bibitem{Masiero:1982fe}
A.~Masiero, D.~V.~Nanopoulos, K.~Tamvakis and T.~Yanagida,
``Naturally Massless Higgs Doublets in Supersymmetric SU(5),''
Phys. Lett. B \textbf{115} (1982), 380-384;
%doi:10.1016/0370-2693(82)90522-6
%398 citations counted in INSPIRE as of 21 Aug 2020
%\cite{Grinstein:1982um}
%\bibitem{Grinstein:1982um}
B.~Grinstein,
``A Supersymmetric SU(5) Gauge Theory with No Gauge Hierarchy Problem,''
Nucl. Phys. B \textbf{206} (1982), 387.
%doi:10.1016/0550-3213(82)90275-9
%348 citations counted in INSPIRE as of 21 Aug 2020

%\cite{Ellis:2002vk}
\bibitem{Ellis:2002vk}
J.~R.~Ellis, D.~V.~Nanopoulos and J.~Walker,
``Flipping SU(5) out of trouble,''
Phys. Lett. B \textbf{550} (2002), 99-107
%doi:10.1016/S0370-2693(02)02956-8
[arXiv:hep-ph/0205336 [hep-ph]].
%67 citations counted in INSPIRE as of 27 Sep 2020

%\cite{Dorsner:2004xx}
\bibitem{Dorsner:2004xx}
I.~Dorsner and P.~Fileviez Perez,
``Distinguishing between SU(5) and flipped SU(5),''
Phys. Lett. B \textbf{605} (2005), 391-398
%doi:10.1016/j.physletb.2004.11.051
[arXiv:hep-ph/0409095 [hep-ph]].
%22 citations counted in INSPIRE as of 27 Sep 2020

%\cite{Kyae:2005nv}
\bibitem{Kyae:2005nv}
B.~Kyae and Q.~Shafi,
``Flipped SU(5) predicts delta T/T,''
Phys. Lett. B \textbf{635} (2006), 247-252
%doi:10.1016/j.physletb.2006.03.007
[arXiv:hep-ph/0510105 [hep-ph]].
%53 citations counted in INSPIRE as of 19 Aug 2020
%
%\cite{Ellis:2017jcp}
\bibitem{Ellis:2017jcp}
J.~Ellis, M.~A.~G.~Garcia, N.~Nagata, D.~V.~Nanopoulos and K.~A.~Olive,
``Starobinsky-like Inflation, Supercosmology and Neutrino Masses in No-Scale Flipped SU(5),''
JCAP \textbf{07} (2017), 006
%doi:10.1088/1475-7516/2017/07/006
[arXiv:1704.07331 [hep-ph]].
%21 citations counted in INSPIRE as of 21 Aug 2020
%
%\cite{Ellis:2018moe}
\bibitem{Ellis:2018moe}
J.~Ellis, M.~A.~G.~Garcia, N.~Nagata, D.~V.~Nanopoulos and K.~A.~Olive,
``Symmetry Breaking and Reheating after Inflation in No-Scale Flipped SU(5),''
JCAP \textbf{04}, 009 (2019)
%doi:10.1088/1475-7516/2019/04/009
[arXiv:1812.08184 [hep-ph]].
%15 citations counted in INSPIRE as of 01 Dec 2020
%
%\cite{Ellis:2019opr}
\bibitem{Ellis:2019opr}
J.~Ellis, M.~A.~G.~Garcia, N.~Nagata, D.~V.~Nanopoulos and K.~A.~Olive,
``Superstring-Inspired Particle Cosmology: Inflation, Neutrino Masses, Leptogenesis, Dark Matter \textbackslash{}\& the SUSY Scale,''
JCAP \textbf{01}, 035 (2020)
%doi:10.1088/1475-7516/2020/01/035
[arXiv:1910.11755 [hep-ph]].
%9 citations counted in INSPIRE as of 01 Dec 2020

%\cite{Barbieri:1982eh}
\bibitem{Barbieri:1982eh}
R.~Barbieri, S.~Ferrara and C.~A.~Savoy,
``Gauge Models with Spontaneously Broken Local Supersymmetry,''
Phys. Lett. B \textbf{119} (1982), 343;
%doi:10.1016/0370-2693(82)90685-2
%1797 citations counted in INSPIRE as of 21 Aug 2020;
%\cite{Chamseddine:1982jx}
%\bibitem{Chamseddine:1982jx}
A.~H.~Chamseddine, R.~L.~Arnowitt and P.~Nath,
``Locally Supersymmetric Grand Unification,''
Phys. Rev. Lett. \textbf{49} (1982), 970;
%doi:10.1103/PhysRevLett.49.970
%2086 citations counted in INSPIRE as of 21 Aug 2020;
%\cite{Nilles:1982dy}
%\bibitem{Nilles:1982dy}
H.~P.~Nilles, M.~Srednicki and D.~Wyler,
``Weak Interaction Breakdown Induced by Supergravity,''
Phys. Lett. B \textbf{120} (1983), 346
%doi:10.1016/0370-2693(83)90460-4
%741 citations counted in INSPIRE as of 21 Aug 2020;
%\cite{Hall:1983iz}
%\bibitem{Hall:1983iz}
L.~J.~Hall, J.~D.~Lykken and S.~Weinberg,
``Supergravity as the Messenger of Supersymmetry Breaking,''
Phys. Rev. D \textbf{27} (1983), 2359-2378
%doi:10.1103/PhysRevD.27.2359
%1563 citations counted in INSPIRE as of 21 Aug 2020
%

%\cite{Lazarides:1996dv}
\bibitem{Lazarides:1996dv}
G.~Lazarides, R.~K.~Schaefer and Q.~Shafi,
``Supersymmetric inflation with constraints on superheavy neutrino masses,''
Phys. Rev. D \textbf{56} (1997), 1324-1327
%doi:10.1103/PhysRevD.56.1324
[arXiv:hep-ph/9608256 [hep-ph]].
%160 citations counted in INSPIRE as of 29 Aug 2020

%\cite{Lazarides:2001zd}
\bibitem{Lazarides:2001zd}
G.~Lazarides,
``Inflationary cosmology,''
Lect. Notes Phys. \textbf{592} (2002), 351-391
[arXiv:hep-ph/0111328 [hep-ph]].
%85 citations counted in INSPIRE as of 29 Aug 2020

%\cite{Ellis:1984eq}
\bibitem{Ellis:1984eq}
J.~R.~Ellis, J.~E.~Kim and D.~V.~Nanopoulos,
``Cosmological Gravitino Regeneration and Decay,''
Phys. Lett. B \textbf{145} (1984), 181-186
doi:10.1016/0370-2693(84)90334-4
%851 citations counted in INSPIRE as of 11 Sep 2020
%

%\cite{Ellis:1984er}
\bibitem{Ellis:1984er}
J.~R.~Ellis, D.~V.~Nanopoulos and S.~Sarkar,
``The Cosmology of Decaying Gravitinos,''
Nucl. Phys. B \textbf{259} (1985), 175-188
doi:10.1016/0550-3213(85)90306-2
%442 citations counted in INSPIRE as of 11 Sep 2020

%\cite{Moroi:1993mb}
\bibitem{Moroi:1993mb}
T.~Moroi, H.~Murayama and M.~Yamaguchi,
``Cosmological constraints on the light stable gravitino,''
Phys. Lett. B \textbf{303} (1993), 289-294
doi:10.1016/0370-2693(93)91434-O
%556 citations counted in INSPIRE as of 11 Sep 2020

%\cite{Kawasaki:2004yh}
\bibitem{Kawasaki:2004yh}
M.~Kawasaki, K.~Kohri and T.~Moroi,
``Hadronic decay of late - decaying particles and Big-Bang Nucleosynthesis,''
Phys. Lett. B \textbf{625} (2005), 7-12
doi:10.1016/j.physletb.2005.08.045
[arXiv:astro-ph/0402490 [astro-ph]].
%464 citations counted in INSPIRE as of 11 Sep 2020

%\cite{Antusch:2010mv}
\bibitem{Antusch:2010mv}
S.~Antusch, J.~P.~Baumann, V.~F.~Domcke and P.~M.~Kostka,
``Sneutrino Hybrid Inflation and Nonthermal Leptogenesis,''
JCAP \textbf{10} (2010), 006
doi:10.1088/1475-7516/2010/10/006
[arXiv:1007.0708 [hep-ph]].
%32 citations counted in INSPIRE as of 11 Sep 2020

%\cite{deVries:2015hva}
\bibitem{deVries:2015hva}
K.~J.~de Vries, E.~A.~Bagnaschi, O.~Buchmueller, R.~Cavanaugh, M.~Citron, A.~De Roeck, M.~J.~Dolan, J.~R.~Ellis, H.~Flächer, S.~Heinemeyer, G.~Isidori, S.~Malik, J.~Marrouche, D.~Martinez Santos, K.~A.~Olive, K.~Sakurai and G.~Weiglein,
``The pMSSM10 after LHC Run 1,''
Eur. Phys. J. C \textbf{75} (2015) no.9, 422
%doi:10.1140/epjc/s10052-015-3599-y
[arXiv:1504.03260 [hep-ph]].
%92 citations counted in INSPIRE as of 13 Sep 2020














\end{thebibliography}
\end{document}